\documentclass[a4paper,fleqn]{cas-dc}

\usepackage[switch]{lineno}
\usepackage[authoryear]{natbib}
\usepackage{gensymb}
\usepackage{comment}
\usepackage{caption}
\usepackage{setspace}
\usepackage{booktabs}
\usepackage{multirow}
\usepackage{soul}

\RenewDocumentCommand{\printorcid}{}{}
\newcommand{\U}[1]{\ensuremath{\mathrm{#1}}}
\newcommand{\sav}[1]{\left\langle {\smash{#1}} \right\rangle}
\renewcommand{\vec}[1]{\boldsymbol{#1}}
\renewcommand{\d}{\textnormal{d}}
\renewcommand{\perp}{h}
\newcommand{\review}[1]{{\color{black}#1}}

\def\tsc#1{\csdef{#1}{\textsc{\lowercase{#1}}\xspace}}
\tsc{WGM}
\tsc{QE}
\tsc{EP}
\tsc{PMS}
\tsc{BEC}
\tsc{DE}

\begin{document}
\let\WriteBookmarks\relax
\def\floatpagepagefraction{1}
\def\textpagefraction{.001}

\shorttitle{Journal of Wind Engineering and Industrial Aerodynamics}

\shortauthors{Huang et~al.}


\title [mode = title]{Building drag and shielding in a realistic urban environment}

\author[1]{Jingzi Huang}
\cormark[1]
\ead{jingzi.huang17@imperial.ac.uk}
\author[2]{Omduth Coceal}
\author[3]{Marco Placidi}
\author[4]{Zheng-Tong Xie}
\author[1]{Maarten {van Reeuwijk}}

\affiliation[1]{organization={Department of Civil and Environmental Engineering, Imperial College London}, country={UK}}
\affiliation[2]{organization={Department of Meteorology, University of Reading}, country={UK}}
\affiliation[3]{organization={EnFlo Laboratory, School of Engineering, University of Surrey}, country={UK}}
\affiliation[4]{organization={Department of Aeronautical and Astronautical Engineering, University of Southampton}, country={UK}}

\begin{abstract}
\review{Shielding by upstream buildings is a fundamental control on urban drag, yet its influence remains poorly quantified in realistic urban environments. Here, we investigate shielding effects using building-resolved large-eddy simulations of the University of Bristol campus, comprising 110 buildings of varying height, shape and orientation. Twenty-four wind directions are considered, allowing each building to experience a wide range of upstream shielding conditions. While the total drag of the campus exhibits only moderate directional variability, the drag acting on individual buildings varies substantially. In the present case, approximately $20\%$ of buildings account for $80\%$ of the total drag, which is primarily attributed to a small number of large buildings that contribute disproportionately high drag forces. To quantify shielding, we introduce two dimensionless parameters: the upstream fetch ratio, $L_s/H_s$, and the relative height ratio, $H_s/H$, where $L_s$ is the distance to the nearest upstream obstacle, $H_s$ is the height of the upstream obstacle, and $H$ is the height of the target building. These parameters distinguish between near- and far-wake conditions and between sheltered and exposed buildings, providing a simple method to characterise shielding effects in realistic urban environments. The study provides valuable quantitative insight into drag and shielding in the Bristol campus morphology; more importantly, it establishes a general framework for analysing drag and shielding that can be applied in other complex urban environments. The results identify shielding as a primary control on building drag and motivate shielding-aware measures of effective frontal area and drag coefficient.}
\end{abstract}

\begin{keywords}
Realistic urban canopy \sep Drag coefficient \sep Wind direction \sep Shielding effect  \sep Large-Eddy Simulation
\end{keywords}

\maketitle

\section{Introduction}
Aerodynamic drag exerted on urban buildings plays a pivotal role in shaping urban boundary layers, particularly in modern cities characterised by clusters of high-rise buildings, dense layouts, and heterogeneous spatial arrangements \citep{Oke2017}. Substantial efforts have been devoted to investigating aerodynamic drag across a variety of urban forms for wind engineering applications, and to parameterising it within urban canopy models (UCMs) used in mesoscale weather forecasting models, such as the UK Unified Model (UM) and the US Weather Research and Forecasting (WRF) model. Early studies typically inferred the drag from bulk aerodynamic properties of entire districts, such as the roughness length $z_0$ and displacement height $z_d$ \citep{Rotach1993, Grimmond1999, Roth2000}, or through morphological indicators such as plan-area index $\lambda_p$ and frontal-area index $\lambda_f$ \citep{Cheng2002, Barlow2002, Hagishima2009, Placidi2015}. More recent studies, benefiting from advances in computational fluid dynamics (CFD), including direct numerical simulation (DNS) and large-eddy simulation (LES), have enabled building-resolved simulations down to the scale of small turbulent structures. As a result, aerodynamic drag can now be estimated at much higher resolution, for example through distributed drag formulations that account for vertical variations within the canopy region \citep{Coceal2006, Xie2006, Dejoan2010, Leonardi2010, Sutzl2020}, as well as volumetric drag coefficients associated with the local plan-area index $\lambda_p^*$ and frontal-area index $\lambda_f^*$ within localised units \citep{Zhang2025, MVR2025}.

The interaction between the urban boundary layer and buildings generates more complex turbulent flow structures, for example, by enhancing variability in wind speed \citep{Xie2008, MacGarry2025} and direction \citep{xie2009large}. Tall buildings \citep{Xie2008,fuka2018scalar}, especially when arranged in clusters \citep{wang2026flow}, generate substantial aerodynamic drag, leading to significant reductions in local wind speed \citep{Mishra2023} and complex non-linear dynamics in the turbulence \citep{Mishra2024}. Street canyons, for instance, impose geometric confinement that constrains the wind to align preferentially with the street axis. Variations in building height further modulate wind direction by inducing secondary circulations and enhancing vertical turbulent exchange \citep{Hang2012}. In addition, irregular building morphologies promote flow separation and corner accelerations, leading to rapid and spatially heterogeneous directional fluctuations over short distances \citep{Moonen2012}.

\review{
Shielding effects, as a specific manifestation of building interaction, constitute an important factor governing building drag. Unlike the frontal area, which is an intrinsic geometric property of a building, shielding effects are associated with extrinsic properties such as the spatial configuration of upstream buildings and incident wind direction \citep{Lam2008, Yu2015, Kim2015, Yan2016, Placidi2017}. In dense urban environments, a building is almost always influenced by shielding, which usually reduces the drag acting on it.
Previous studies have provided important insights into shielding as a key part of building interference effects. The resulting drag force reduction can be substantial, reaching approximately $30\%$--$60\%$ relative to isolated conditions \citep{English1990,  Lam2011, ASNZS1170.2}. \citet{Khanduri1998} defined a shielding factor as the ratio of the aerodynamic force (i.e., drag) on a building under shielding effects to the force on that building in an isolated condition. Considerable efforts have been devoted to investigating interference effects between twin tall buildings \citep{Tang2004, Xie2007, Lam2011, Yu2015, Liang2020}. These studies have shown that drag reduction and pressure redistribution depend strongly on building spacing and relative height. \citet{Kim2015, Yu2015} additionally considered shielding under various wind directions. However, the previous studies mentioned above have largely been restricted to pairs of idealised cuboid buildings arranged in tandem, side-by-side, or staggered configurations, and such simplified configurations do not capture the complexity of realistic urban environments. Although \citet{Hui2012, Yan2016} attempted to consider more complex building configurations, these remained limited and did not reflect the full complexity of real urban environments. Therefore, a systematic characterisation of building drag and shielding effects in realistic urban environments is yet to be investigated.
}
A building that is well exposed under one wind direction may become more sheltered under another; wind direction thus has a substantial influence on shielding and drag, as it not only alters the frontal area of the building facing the wind but, importantly, alters the shielding conditions imposed by upstream structures. However, this aspect also remains insufficiently studied.

\review{The aim of the present study is to characterise shielding effects in realistic urban environments through building drag.} We analyse turbulent flow through a realistic representation of the University of Bristol campus (UK), comprising 110 buildings of substantially varying shapes and orientations, using building-resolved LES under 24 wind directions. This dataset enables a building-by-building assessment of how drag is modified by upstream shielding across a wide range of local geometric configurations. Based on this analysis, we identify controlling geometric parameters for the shielding effects, classify the resulting drag behaviour into distinct regimes, and examine how shielding influences the characterisation of drag coefficients in realistic urban settings. The manuscript is organised as follows: \S \ref{sec: Methodology} introduces the methodology for computing drag and drag coefficients; \S \ref{sec: simulation details} describes the LES setup and implementation details; \S \ref{sec: results} presents the results and discussion; and the main conclusions are summarised in \S \ref{sec: conclusions}.

\section{Methodology} \label{sec: Methodology}
To analyse building drag and upstream shielding, this section primarily formulates the streamwise drag at a building-resolved scale based on the algorithm from \citet{MVR2025}. Subsequently, the total drag and bulk drag coefficient of an individual building are defined. It is worth clarifying that the term `streamwise' always refers to the prescribed upstream wind direction, while `spanwise' denotes the direction perpendicular to it. The wind can also induce a so-called `across-wind' force (or `lift force', `lateral force') in the spanwise direction \citep{Claus2012}, so that the resultant force acting on buildings is not necessarily aligned with the prescribed inflow direction. \review{The deflection of the wind velocity can be locally strong within the canopy due to wind veering; however, for the plane-averaged velocity, it is typically less than $7\degree$, indicating that the induced force remains dominated by the streamwise component \citep{Claus2012}. The present study focuses primarily on the drag in the streamwise direction; the across-wind force is discussed only briefly in \S~\ref{sec: lift} and is otherwise beyond the main scope of this work.}

\subsection{Building-resolved Reynolds-averaged drag in the streamwise direction} \label{sec: Reynolds-averaged drag}
\begin{figure*}
    \centering
    \includegraphics[width=12cm]{ 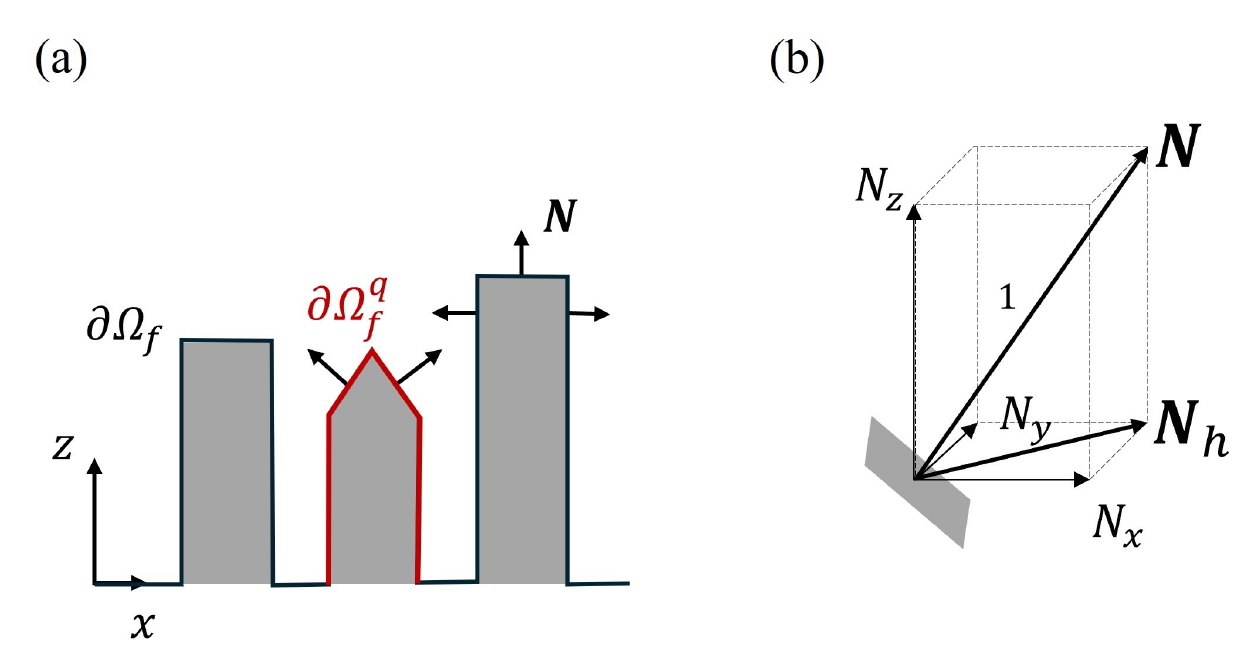}
    \caption{Schematic definition of (a) the solid-fluid interface $\partial \Omega_f$ and the 3-D normal vectors $\vec N$ of the interface, where the solid domain is shown in grey and the fluid domain $\Omega_f$ in white. The surface of an individual building $q$ is outlined in red; (b) the decomposition of $\vec N$, where $\vec N$ is a unit vector pointing into the fluid domain.}
    \label{fig:sketch}
\end{figure*}
Since we focus on the drag in a time-averaged sense and neglect buoyancy effects, the Reynolds-averaged momentum equation in the streamwise direction for incompressible flow over the canopy region can be written as:
\begin{equation}  \label{eq:URANS}
  \frac{\partial \overline u_j \overline u}{\partial x_j} +  \frac{\partial \overline{u_j' u'}}{\partial x_j}
  +\frac{\partial \overline p}{\partial x} 
  = f+\nu \frac{\partial^2 \overline{u}}{\partial x_j^2}  \, .
\end{equation}
Here, we define the coordinate vector $\vec{x} = [x, y, z]^T$, where $x$, $y$, and $z$ denote the streamwise, spanwise, and vertical directions, respectively, and the velocity vector $\vec{u}(\vec{x}) = [u, v, w]^T$, where $u$, $v$, and $w$ are the corresponding velocity components. $p(\vec x)$ is the kinematic deviatoric pressure including density, $f(\vec x)$ is the external force, for example, the large-scale pressure forcing imposed to drive the wind, $\nu$ is the turbulent kinematic viscosity, and the symbol $\overline{\cdot}$ represents the time average. We define the superficial horizontal plane average as:
\begin{equation} \label{eq: plav}
    \sav{\chi}(z) = \frac{1}{A_T} \int_{\Omega_f(z)} \chi \d \vec x_h \, ,
\end{equation}
where $\chi$ is an arbitrary quantity, $A_T$ is the horizontal plane area, $\vec x_h = [x, y]^T$ represents the horizontal vector, and $\Omega_f$ denotes the fluid domain.
Under this averaging operation, the horizontal momentum-transport terms vanish, and Eq. \eqref{eq:URANS} becomes:
\begin{equation} \label{eq:planav_mom}
-\frac{\d \tau_u}{\d z} =  \sav{f}-f_D  \, ,
\end{equation}
\review{
where $\tau_u$ is the total vertical flux of streamwise (kinematic) momentum:
\begin{equation} \label{eq: tau_u}
    \tau_u \equiv \sav{\overline{w} \, \overline u} 
+\sav{\overline{w^\prime u^\prime}}-\nu \frac{\d \sav{\overline{u}}}{\d z} \, .
\end{equation}
Then, we use the triple decomposition framework \citep{Raupach1982, Finnigan2000b}, where the time average is decomposed into a spatial mean and a spatial variation, e.g., $\overline{u} \equiv \sav{\overline{u}}^f + \overline{u}^{\prime\prime}$. The notation $\sav{\chi}^f$ denotes the intrinsic average, i.e.\ the plane average over the fluid field, which can be obtained via Eq. \eqref{eq: plav} by replacing $A_T$ with the area of the fluid region $\Omega_f.$ Applying this decomposition gives:
\begin{equation} \label{eq: uw}
\sav{\overline{w}\,\overline{u}} = \sav{\sav{\overline{w}}^f \sav{\overline{u}}^f} +
\sav{\sav{\overline{w}}^f \overline{u}^{\prime\prime}} + \sav{\overline{w}^{\prime\prime} \sav{\overline{u}}^f}
+ \sav{\overline{w}^{\prime\prime}\overline{u}^{\prime\prime}} \, .
\end{equation}
Note that: first, due to the spatial deviation definition, $\sav{\overline{u}^{\prime\prime}}$ and $\sav{\overline{w}^{\prime\prime}}$ are zero; second, the intrinsic and horizontal averages $\sav{\overline{w}}^f$ and $\sav{\overline{u}}^f$ are constants with respect to horizontal position. Additionally, for the periodic boundary conditions, $\sav{\overline{w}}^f = 0$ due to mass continuity. Therefore, only the last term remains on the right-hand side of Eq.~\eqref{eq: uw}, so that Eq.~\eqref{eq: tau_u} becomes:
\begin{equation} \label{eq:tau_x}
    \tau_u  =  \sav{\overline{w}^{\prime\prime} \, \overline{u} ^{\prime\prime}} +\sav{\overline{w^\prime u^\prime}}-\nu \frac{\d \sav{\overline{u}}}{\d z} \, .
\end{equation} 
The term $\sav{\overline{w}^{\prime\prime} \, \overline{u} ^{\prime\prime}}$ represents the dispersive stress associated with vertical momentum transport caused by local variations in the mean velocity. The term $\sav{\overline{w^\prime u^\prime}}$ is the superficial plane-averaged turbulent flux, and the last term is the momentum flux associated with the viscosity, which can be negligible for turbulent flows.
}

The term $f_D$ is the plane-averaged drag in the streamwise direction that results from performing the plane integral of the pressure and viscous terms in Eq. \eqref{eq:URANS}, and is given by \citep{MVR2025}:
\begin{equation} \label{eq:fD}
    f_{D}(z) =  - \frac{1}{A_T} \oint_{\partial \Omega_f(z)} \, \left( \overline p \frac{  N_x}{|\vec N_\perp|} - \nu \frac{\partial \overline u}{\partial x_j}  \frac{ N_j}{|\vec N_\perp|} \right)\d s \, .
\end{equation}
This is a line integral of the pressure and viscous stresses along the solid surface $\partial \Omega_f(z)$. \review{Here, $\partial \Omega_f(z)$ denotes the 1-D closed contour obtained by intersecting the solid surface $\partial \Omega_f$ with the horizontal plane at height $z$. This contour integral yields the distributed drag $f_D (z)$ as a function of $z$,} which includes the form drag (component associated with pressure) and the skin drag (component associated with viscosity).

In the equation above, $\vec{N}_h$ represents the horizontal component of the local surface normal. In Fig. \ref{fig:sketch}(a), the black line outlines the solid surface which is the interface between the solid and fluid phases, $\vec N = [\vec N_{\perp}, N_z]^T$ is the normal of the solid surface with $\vec N_{\perp} = [N_x, N_y]^T$ (Fig. \ref{fig:sketch}b), so that $|\vec N_{\perp}|$ is the projected length of the normal on the horizontal plane.

The distributed drag $f_D$ can be obtained via two approaches: 1) indirectly from the budget closure of Eq. \eqref{eq:planav_mom}; and 2) directly from Eq. \eqref{eq:fD}. Equation \eqref{eq:fD} is the preferred approach for three reasons. First, it allows for a direct check on whether the budget closes. Second, it allows for the calculation of components (e.g., the form drag associated with pressure, and the skin drag associated with viscosity). Third, it allows us to calculate the drag of an individual building by restricting the integration domain $\partial \Omega_f$ to a specific building surface $\partial \Omega_{f}^{(q)}$ (see Fig. \ref{fig:sketch}a  for an example outlined in red); here the superscript $(q)$ denotes the building ID. Denoting $Q$ as the entire building group, the solid surface $\partial \Omega_f$ consists of every single building surface and the ground surface $\partial \Omega_{f}^{g}$.
%
%
Therefore, the distributed drag $f_D$ can be split as:
\begin{equation} \label{eq:FDsubsurface}
 f_D = f_D^g + \sum_{q \in Q} f_D^{(q)} \, ,
\end{equation}
where the drag from the buildings is:
\begin{equation} \label{eq:fDq}
    f_D^{(q)}(z) \approx  - \frac{1}{A_T} \oint_{\partial \Omega_{f}^{(q)}(z)} \,  \overline p \frac{  N_x}{|\vec N_\perp|} \, \d s \, .
\end{equation}
%
%
Here, the skin drag is ignored as the form drag is the dominant component on buildings. The skin drag is the only contribution at the ground surface:
\begin{equation}
    f_D^{g}(z) =   \frac{1}{A_T} \oint_{\partial \Omega_{f}^{g}(z)} \,  \nu \frac{\partial \overline u}{\partial x_j}  \frac{ N_j}{|\vec N_\perp|} \d s \, .
\end{equation}
This term is singular, since $|\vec N_\perp| = 0$ for horizontal surfaces. This is to be expected, as a finite amount of momentum is exchanged over an infinitesimal distance $\d z$, i.e. $f_D^{g}(z) = \tau_D^g \delta(z)$, see \cite{Maarten_2021, MVR2025}. Note
that the presence of these apparent singularities is an unavoidable consequence of using a
planar average.
%
The implementation details for computing the surface integrals are described in \S \ref{sec: simulation details} and Appendix \ref{sec: volumetric density}.

\subsection{Vertically cumulative drag}
Integrating the distributed drag along the vertical direction yields the vertically cumulative drag, which represents the total drag that acts on the buildings. In this sense, the plane-averaged total drag stress acting on the entire canopy region can be expressed as:
\begin{equation} \label{eq:tau_D}
    \tau_{D} =  \int_0^{h} f_{D}(z) \d z \, .
\end{equation}
where $h$ is the top of the domain. Note that, above the canopy, the drag is zero by definition because no solid surfaces are present. The quantity $\tau_D$ is also referred to as total kinematic surface shear stress \citep{pope_2000}, and is usually denoted as $\tau_0$.
Similar to Eq. \eqref{eq:FDsubsurface}, $\tau_D$ can be decomposed as
\begin{equation} \label{eq: tau_Dq}
 \tau_{D} = \sum_{q \in Q} \tau_{D}^{(q)} +  \tau_s\, ,
\end{equation}
where $\tau_{D}^{(q)}$ is the drag stress on an individual building:
\begin{equation}
    \tau_{D}^{(q)} =  \int_0^{h} f_{D}^{(q)}(z) \d z \, ,
\end{equation}
and $\tau_s$ is the ground surface stress.
\begin{equation}
    \tau_s =  \int_0^{h} f_{D}^{g}(z) \d z \, ,
\end{equation}
respectively.

Alternatively, we can also use an actual drag force to incorporate the influence of plane area, e.g., the actual drag acting on the entire canopy $F_D = A_T \tau_D$, and the actual force acting on a single building $q$ is $F_D^{(q)} = A_T \tau_{D}^{(q)}$.

\subsection{Bulk drag coefficient}
The drag of the entire domain can be evaluated via a single bulk drag coefficient $c_D$ \citep{Coceal2004, Sutzl2021}:
\begin{equation} \label{eq: Cd}
    \tau_D = \frac{1}{2} \rho_0 \lambda_f c_D U^2 \, ,
\end{equation}
or equivalently
\begin{equation} \label{eq: F_D}
    F_D = \frac{1}{2}\rho_0 A_f c_D U^2 \, ,
\end{equation}
where $\rho_0$ is the density of air, $\lambda_f$ and $A_f$ are the frontal-area index and the frontal area of the buildings, respectively, $A_f= \lambda_f A_T$, and $U$ is a characteristic velocity at a reference height. Note that the frontal area is associated with the streamwise direction.

Similarly, the bulk drag coefficient $C_D^{(q)}$ of an individual building $q$ is defined via
%
%
%
\begin{equation} \label{eq: F_D^q}
  F_{D}^{(q)} = \frac{1}{2}\rho_0 A_{f}^{(q)} C_{D}^{(q)} U^2 \, ,
\end{equation}
where $\lambda_{f}^{(q)}$ and $A_{f}^{(q)}$ are the frontal-area index and frontal area of building $q$, respectively, with $A_{f}^{(q)} = A_T \lambda_{f}^{(q)}$. 
\review{
The characteristic velocity $U$ is typically taken at a specific height \citep{Grimmond1999, Hagishima2009}. In this study, it is taken as the plane-averaged streamwise velocity at $z = 30$ m, i.e., $U_{30}$, for two reasons. First, this height lies approximately $2.5 h_m$ above the ground where the mean flow is less disturbed by the buildings, and is furthermore consistent with the reference height used in previous studies \citep[e.g.,][]{GarciaSanchez2018}. Second, the wind velocity at this height can be readily obtained from mesoscale numerical weather prediction models \citep{Skamarock2008}, facilitating the practical application of the drag coefficient to larger-scale simulations. A sensitivity study reported by \citet{Claus2010} has confirmed that the choice of reference height primarily affects the magnitude of the drag coefficient while leaving the directional variation and relative trends essentially unchanged.
}

\review{
\subsection{Effective drag coefficient}
The drag coefficient, $C_D$, is commonly used as a dimensionless measure of the intrinsic aerodynamic resistance of a building. For example, it is typically measured for an isolated building in a wind tunnel, where no upstream shielding is present, as in studies on the CAARC buildings \citep[e.g.][]{Melaku2024}. In real urban environments, however, buildings are rarely isolated. Upstream obstacles modify the approaching flow, and consequently, the drag force acting on a building inherently incorporates the influence of shielding. Drag coefficients evaluated directly from such drag forces should therefore be interpreted as \emph{effective} drag coefficients, denoted here as $C_{D,\mathrm{eff}}$, to reflect the shielding effects embedded within them. Shielding is unavoidable in dense layouts such as the one examined here; hereafter, we therefore calculate and examine the effective drag coefficient $C_{D,\mathrm{eff}}$:
\begin{equation} \label{eq: F_D^q_e}
  F_{D}^{(q)} = \frac{1}{2}\rho_0 A_{f}^{(q)} C_{D,\mathrm{eff}}^{(q)} U^2 \, .
\end{equation}
%


%
%
However, for brevity, the qualifier `effective' is omitted in the narrative until \S \ref{sec: excluding shielding effect}, with the distinction preserved through the symbolic notation. In \S \ref{sec: excluding shielding effect}, we analyse the relationship between the effective ($C_{D,\mathrm{eff}}$) and the conventional drag coefficients ($C_D$); the two terms are strictly distinguished therein.

}

\section{Simulation details} \label{sec: simulation details}
\begin{figure*}
    \centering
    \includegraphics[width=16cm]{ 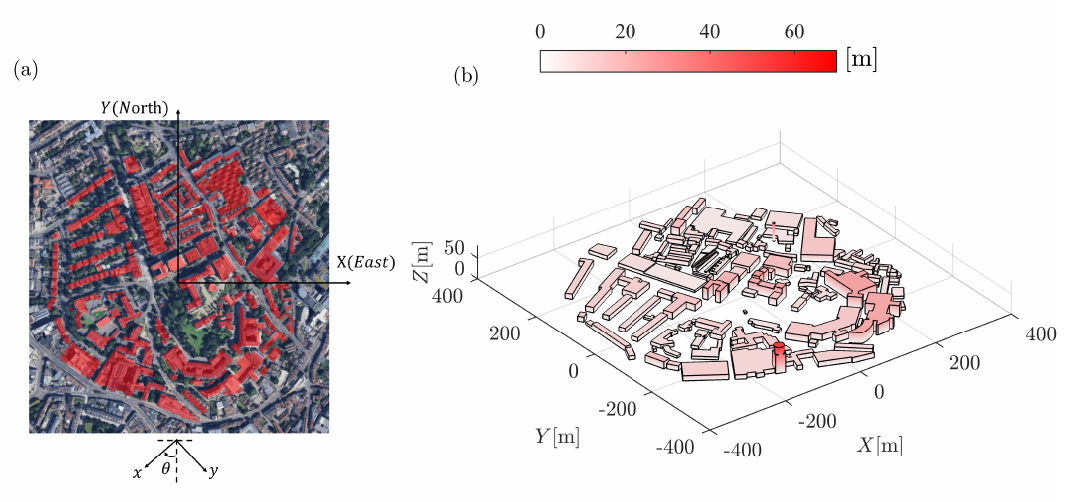}
    \caption{(a) A satellite plan view of the campus of the University of Bristol, overlaid with a footprint of the simulation morphology. From Google Maps. (b) A 3-D view of the simulated campus morphology \citep{Bi2025}, with the colour indicating the building height level. A global Cartesian coordinate system $(X,Y,Z)$ is defined with the origin at the campus centre of the plane view, where positive $X, Y$ point eastward and northward, respectively. A local coordinate system $(x, y, z)$ is also defined, where positive $x, y$ are along the wind streamwise and spanwise directions, respectively, while vertical $z$ coincides with $Z$. The wind direction forms an angle $\theta$, defined as the clockwise rotation of the positive local $x$-axis from the North.}
    \label{fig: Fig1}
\end{figure*}
Figure \ref{fig: Fig1} presents both the plan view and 3-D view of the simulated domain --- the campus of the University of Bristol. The domain contains 110 individual buildings of varying shapes and orientations, with a mean full-scale building height of $h_m = 12$ m, and a maximum height of $h_{\max} = 63$ m, corresponding to the Wills Memorial Tower, which has a dark red top in Fig. \ref{fig: Fig1}(b). As indicated in Fig. \ref{fig: Fig1}(b), the campus buildings exhibit a circular layout around the campus centre, and a large horizontal extent but relatively limited vertical development: the plan-area index and frontal area index of the morphology are $\lambda_p = 0.24$ and $\lambda_f = 0.12$ (for a south wind $\theta = 180\degree$), respectively. Note that the building layout is based on a wind tunnel model \citep{Bi2025}, where the buildings are closely circled around the campus centre, forming a dense layout and providing substantial shielding effects --- we shall see this in fetch distance and shielding height in \S \ref{sec: two parameters}. 
Two coordinate systems are established, see Fig. \ref{fig: Fig1}(a), a fixed global one $(X, Y, Z)$ with positive $X, Y$ aligning with East and North, respectively, and a local one $(x, y, z)$ with positive $x, y$ aligning with the streamwise and spanwise directions, respectively. According to \citet{WMO2018}, the wind direction $\theta$ is defined as the direction from which the wind is blowing, measured clockwise from true North, as labelled in Fig. \ref{fig: Fig1}(a). The local and global coordinates coincide at $\theta = 270\degree$. 
Finally, although Bristol City is characterised by strong terrain variations, the topography is not considered in this study. We assume that within this specific campus domain, the local elevation changes are not significant. Therefore, the ground surface is flat in this study.

The urban flows over the campus are simulated using the open-source large-eddy simulation code uDALES \citep{Suter2022, Owens2024}, which implements the solid boundary using the immersed boundary method (IBM). The dynamics near the boundary are parametrised by the logarithmic wall functions \citep{Uno1995, Suter2022}. The eddy viscosity is calculated following Vreman's subgrid model \citep{Vreman2004}. The code employs a second-order central difference scheme on a staggered Arakawa C-grid for spatial discretisation and an explicit third-order Runge-Kutta scheme for time integration.

The simulation domain size is $L_X \times L_Y \times L_Z = 800 \times 800 \times 300 \, \U{m^3}$ with an equidistant grid size $N_X \times N_Y \times N_Z = 400 \times 400 \times 300$. The domain top is free-slip, and the four lateral sides of the domain have a periodic boundary condition. 
\review{The use of periodic boundary conditions is a well-established approach in LES of urban canopy flows \citep{Xie2006, Letzel2008, MVR2025}. The primary objective of this study is to examine the building shielding effect on drag, which is mainly governed by local flow interactions within the canopy. Specifically, the shielding is determined by the relative positioning of upstream and downstream buildings, rather than by the large-scale flow development or wake recovery effects that could be influenced by the artificial periodic boundary conditions. Moreover, the current domain size is nearly two orders of magnitude greater than the turbulence integral length scale, suggesting the impact of artificial repetition of flow and turbulence is not significant.} The simulation is thermally neutral with a constant pressure gradient forcing $\d P/\d x = 1.25 \times 10^{-5}$ kg m$^{-2}$s$^{-2}$ imposed to drive the wind, therefore, the total shear stress $\tau_0$ is predetermined accordingly as
$
    \tau_0 = \frac{1}{\rho_0} \frac{V_{air}}{A_T} \frac{\d P}{\d x} = 3.7 \times 10^{-3} \, \U{m}^2/ \U{s}^2 \, ,    
$, 
where $V_{air}$ is the air volume, which is constant across all simulated cases, and noting that $\tau_0$ is consistent with $\tau_D$.
To account for directional dependence, a total of 24 wind directions are simulated, with the inflow wind direction varying from $\theta = 0 \degree $ (a north wind, aligned with the negative global $Y$-axis) to $\theta = 345 \degree$ in $15 \degree$ increments. For ease of post-processing, instead of rotating the wind direction, the campus geometry is actually rotated in simulations, while the wind direction is kept aligned with the positive $Y$-axis to achieve the same effect. Each simulation runs for $240\, 000 \,\U{s}$, of which the final $168\, 000 \, \U{s}$ are used to obtain converged time-averaged statistics.

\review{uDALES is a well-established LES code that has demonstrated excellent performance in numerous previous studies, including comparisons against wind tunnel experiments \citep{Lim2022, Owens2024, Fellini2026}. A dedicated two-step verification and validation is performed in Appendix \ref{sec: validation}. First, a cross-code comparison is performed between the current uDALES results and simulations conducted with the PALM 6.0 code \citep{Maronga2020}, another widely used LES code that has been validated across many studies \citep{Gronemeier2021, Resler2021, Anders2023, MacGarry2025, wang2026flow}, at the same resolution for four wind directions $\theta = 0^\circ, 90^\circ, 180^\circ, 270^\circ$. Secondly, a grid-sensitivity analysis is performed by comparing the current baseline uDALES results against higher-resolution uDALES simulations (with grid spacing halved) for $\theta = 0^\circ$ and $180^\circ$. Both the cross-code verification and the grid-sensitivity analysis show good agreement, confirming that the present simulation setup and numerical settings are adequate for the study.}

The surface calculation is an important feature of the uDALES. Besides the conventional Cartesian grid, uDALES employs an additional surface mesh defined on building surfaces to support the IBM and surface-related calculations. Building surfaces are discretised into triangular facets, for which geometric information (area and normal vectors), the nearest Cartesian grid cells, and surface variables such as pressure and fluxes are stored. This facet-based representation enables an efficient and accurate treatment of IBM and surface processes, particularly for buildings with irregular geometries. Using the facet pressure together with facet areas and normals, we introduce a volumetric drag density field, denoted as $\rho_D(\vec x)$, which is converted from the facet pressure and is non-zero only in the nearest Cartesian grid cells to the surface facets. The detailed definition of the volumetric density is given in the Appendix \ref{sec: volumetric density}. By construction, the drag force can then be evaluated as $f_D = \sav{\rho_D}$, such that the surface integral in the drag formulation Eq. \eqref{eq:fD} is transformed into a planar integral over the Cartesian grid. This transformation renders the surface calculation more straightforward and flexible within the Cartesian coordinate system.

\section{Results} \label{sec: results}
This section first analyses the horizontally plane-averaged flow statistics, including the drag, then examines the individual-building drag under different shielding conditions, and finally explores the factors associated with the shielding effects.

\subsection{Plane-averaged flow statistics for different directions}
\begin{figure*}
    \centering
    \includegraphics{ 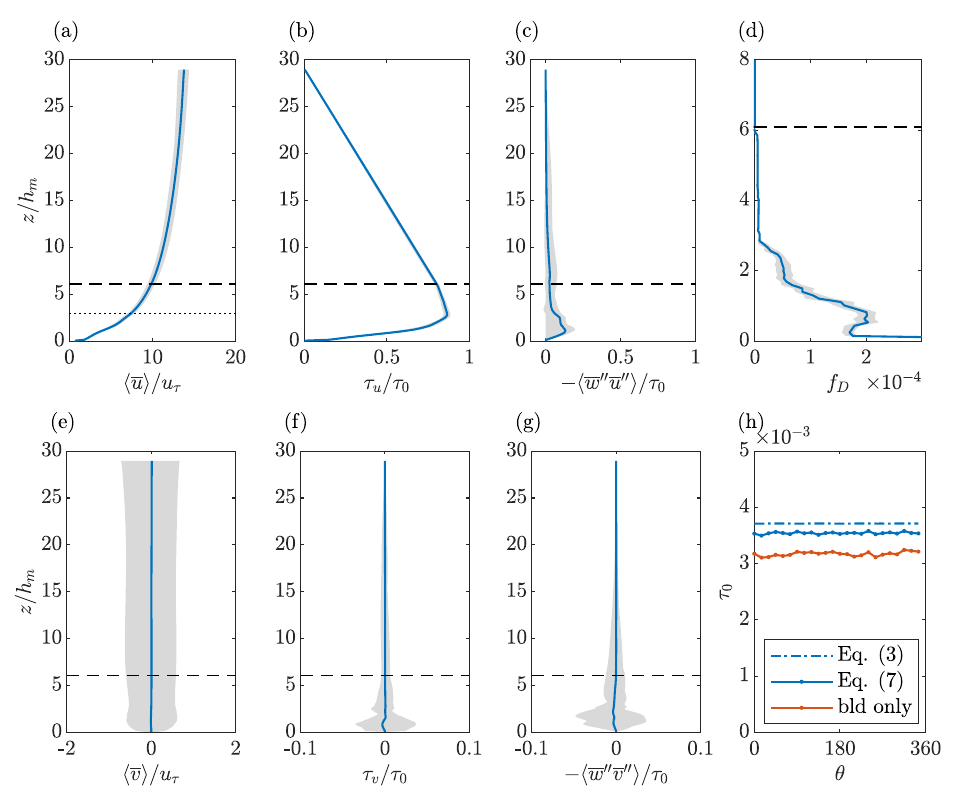}
    \caption{The vertical profiles of plane-averaged velocity (a, e), total kinematic stress (b, f) and dispersive stress (c, g) in the streamwise direction (upper row) and spanwise direction (bottom row), respectively. (d) Streamwise distributed drag $f_D$. The solid line represents the overall mean value across the wind directions, while the shaded band denotes the range between the minimum and maximum values. The dashed horizontal line marks the level of maximum building height $h_{\max}$, while the dotted horizontal line in (a) marks the level of $z = 30$ m. The height is normalised by the mean building height $h_m$. The total shear stress $\tau_0$ from different approaches (h), according to Eq. \eqref{eq:planav_mom} and Eq. \eqref{eq:fD}, respectively, overlaid with the stress only contributed by the building surface. }
    \label{fig: Fig2}
\end{figure*}

Figure \ref{fig: Fig2} summarises the plane-averaged flow statistics over all wind-direction simulations. The solid lines denote the directional mean, while the shaded bands indicate the range across wind directions. Overall, the streamwise velocity (Fig.~\ref{fig: Fig2}a) shows only moderate directional variability, especially within the lower canopy where the campus morphology is relatively symmetric, but the directional variation increases with height as fewer buildings remain. While the streamwise dispersive stress (Fig.~\ref{fig: Fig2}c) and distributed drag (Fig.~\ref{fig: Fig2}d) show more directional variation in the lower canopy, where most buildings are located, they gradually decrease as the buildings are absent. This is because the buildings are the primary source of the dispersive stress and drag.

The streamwise total kinematic stress $\tau_u/\tau_0$ (Fig.~\ref{fig: Fig2}b) remains nearly identical across wind directions and exhibits a linear dependence above the canopy. This is as expected according to Eq. \eqref{eq:planav_mom} under the fixed driving pressure gradient, consistent with boundary-layer theory. The dispersive stress (Fig.~\ref{fig: Fig2}c) is appreciable within the canopy, reflecting spatial heterogeneity induced by the buildings, but is much smaller than the total kinematic stress; hence, the turbulent stress component is dominant. The spanwise velocity, spanwise stress, and spanwise dispersive stress (Fig.~\ref{fig: Fig2}e, f, g) are all small compared with their streamwise counterparts, confirming that the plane-averaged momentum balance is dominated by the streamwise component.

Figure \ref{fig: Fig2}(d) shows that the distributed drag is concentrated within the canopy and decreases rapidly above the main building-height range. Its vertical integral, shown in Fig. \ref{fig: Fig2}(h), is consistent between the momentum-budget estimate and the direct surface-integral calculation, with discrepancies below $5 \%$. The direct calculation slightly underestimates the prescribed stress, likely because the facet mesh is coarser than the Cartesian grid (with one facet associated with approximately six Cartesian grid cells in the present simulations; see Appendix \ref{sec: volumetric density}). The building-only contribution accounts for most of the total drag, whereas the ground surface contributes approximately $10 \%$; therefore, the following analysis focuses on building drag and shielding.

\begin{figure*}
    \centering
    \includegraphics[width = 16cm]{ 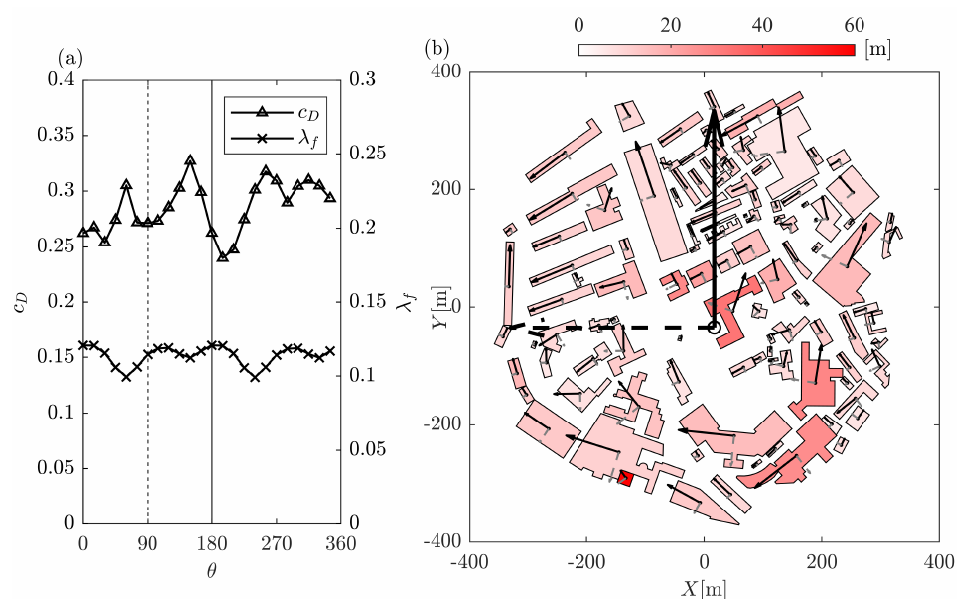}
    \caption{(a) The frontal area index $\lambda_f$ and the drag coefficient $c_D$ of the entire campus varying with the wind direction $\theta$, where the solid and dashed vertical lines label the principal direction and secondary direction of the morphology, respectively. (b) A plane view of the morphology with the principal direction (thick solid) and secondary direction (thick dashed) from the centre and marked by a solid line and a dashed line in (a), respectively. The principal direction and secondary direction of each individual building are also presented with thin arrows.}
    \label{fig: Fig3}
\end{figure*}

Figure \ref{fig: Fig3}(a) shows the variation of the overall frontal area index $\lambda_f$ and the drag coefficient $c_D$ of the campus with respect to wind direction $\theta$. As discussed, to be consistent with the subsequent individual-building study, the ground surface drag is hereafter excluded from the drag coefficient.
The frontal-area index $\lambda_f$ has a mean of $0.12$, and a coefficient of variation of $0.06$ (defined as the
ratio of standard deviation to the mean), while the drag coefficient $c_D$ has a mean of $0.28$, with a coefficient of variation of $0.09$.
Note that the relationship between $c_D$ and $\lambda_f$ is governed by Eq. \eqref{eq: Cd}, in which $U_{30}$ also plays a role.

The principal and secondary directions of the morphology can characterise the spatial orientation and anisotropic heterogeneity of buildings from the geometry, and indicate the wind directions associated with the maximum and minimum building drags, respectively. A height-Weighted Principal Component Analysis (WPCA) is employed \citep{Delchambre2014, Jolliffe2016}, where the eigenvectors of a weighted covariance matrix --- constructed from centred coordinate vectors and the corresponding local building heights --- define the principal and secondary axes of the geometry, see Appendix \ref{sec: WPCA} for more technical details.

Figure \ref{fig: Fig3}(b) presents the principal and secondary directions of the entire campus, shown as bold black arrows, overlaid on the map of local building height. The vectors are centred at $(17\, \U{m}, -35\, \U{m})$, which is close to the geometric centre of the domain $(0,0)$ but is slightly shifted towards the fourth quadrant due to the presence of taller buildings in that region. Although the principal and secondary directions align approximately with the north-south and east-west axes, respectively, their lengths are nearly identical and close to the diameter of the circular campus patch. This indicates a symmetric and approximately isotropic spatial distribution of the campus buildings, consistent with the plan-view observation. As also marked in Fig. \ref{fig: Fig3}(a), the values of $c_D$ and $\lambda_f$ corresponding to the principal and secondary directions are close within $5\%$.

In contrast, individual buildings, whose principal and secondary directions are also marked in Fig. \ref{fig: Fig3}(b), exhibit much stronger directional heterogeneity owing to their distinct shapes. Specifically, for most buildings, it is obvious that, the frontal area is small along the principal direction, whereas it is much larger along the secondary direction, indicating that the drag has a strong wind-direction dependence, if the building is isolated. Overall, the aerodynamic drag of a single building is far more sensitive to wind direction than that of the campus as a whole.

\subsection{The drag force on individual buildings}
\begin{figure*}
    \centering
    \includegraphics[width=16cm]{ 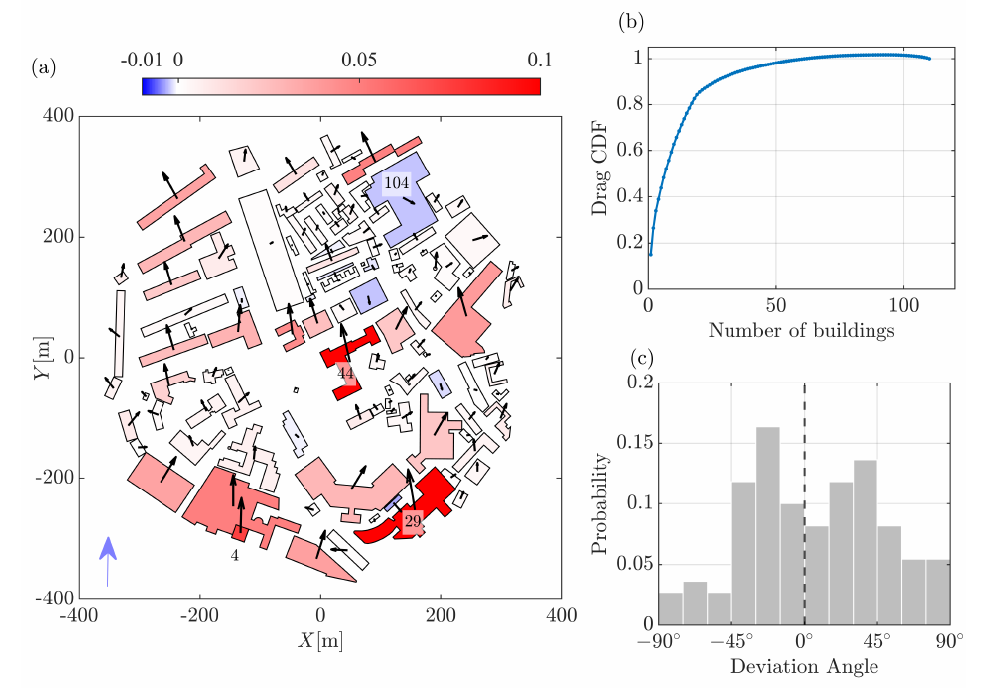}
    \caption{Building drag under wind direction $\theta = 180^\circ$ (wind direction indicated by the light blue arrow at the bottom-left corner): (a) Drag of each building represented by colour, normalised by the total campus drag. The three buildings with the largest drag are numbered. Black arrows indicate the resultant force direction of each building, with lengths proportional to the logarithm of the resultant force magnitude. (b) Cumulative distribution function (CDF) of drag, with buildings ranked in descending order of drag. (c) Probability distribution of the deviation angle between the resultant force directions and the drag direction. The deviation angles range from $-90^\circ$ to $+90^\circ$, where positive values indicate clockwise deviation and negative values indicate anti-clockwise deviation.}
    \label{fig: Fig4}
\end{figure*}
To investigate the drag acting on individual buildings $F_D^{(q)}$, we first show the drag and related statistics under one wind direction, and then examine the drag averaged over all wind directions.
\review{The colour plot in Fig. \ref{fig: Fig4}(a) shows the drag force of each building under wind direction $\theta = 180^\circ$, normalised by the total drag force $F_D$.} It shows that most buildings experience a positive drag (along the wind direction), while a small proportion of buildings experience a slight negative drag (opposite to the wind direction); these buildings are located in the wake region (e.g., the building numbered as `104'). Note that this dimensional drag is affected by the size of the buildings; the 3 buildings experiencing the largest drag (numbered as `29', `44', `4') are either particularly tall or have large footprints.

The values of drag for each building shown in Fig. \ref{fig: Fig4}(a) are then ranked in descending order, and their Cumulative Distribution Function (CDF) is shown in Fig. \ref{fig: Fig4}(b). The CDF demonstrates that a small subset of buildings contributes disproportionately to the total drag. Specifically, the first 6 buildings with the largest drag account for approximately $50\%$ of the total drag, while the first 20 buildings with the largest drag together contribute approximately $80\%$. Conversely, nearly half of the buildings (e.g., the 50 lowest-drag buildings) experience very low drag force in total. The slight exceedance of the CDF above unity in some intervals arises from buildings with small negative drag. This highly uneven distribution highlights the strong spatial heterogeneity across the campus. 

We further conducted a similar analysis on the drag, but averaged over all 24 wind directions, and a similar uneven result was obtained: nearly $20\%$ of the buildings contribute $80\%$ of the total drag. Note that this result depends on the specific campus layout; for example, the campus has a circular and symmetric layout, so that the wind-direction variation is moderate as shown in Fig.~\ref{fig: Fig3}(a). Moreover, the dimensional drag force is associated with the size of the buildings, and there are indeed some large buildings on the campus that contribute more.

\review{
\subsection{The across-wind force on individual buildings} \label{sec: lift}
Although the across-wind force, i.e., spanwise lateral force, is not the primary focus of this study, it is worth briefly investigating its effect before proceeding to the more detailed analysis of drag. Analogous to the drag in Eq. \eqref{eq:fDq}, the plane-averaged across-wind force of building $q$ can be written as
\begin{equation} \label{eq:fLq}
    f_L^{(q)}(z) \approx - \frac{1}{A_T} \oint_{\partial \Omega_{f}^{(q)}(z)} \overline p \frac{N_y}{|\vec N_\perp|} \, \d s \, ,
\end{equation}
where $f_L^{(q)}$ denotes the across-wind force associated with the pressure force acting in the spanwise direction through the $N_y$ component of the building-surface normal. Here, the contributions from the roof and ground surfaces are neglected. The total across-wind force acting on building $q$ and on the campus as a whole is then given by
\begin{equation}
    F_{L}^{(q)} = A_T \int_0^{h} f_{L}^{(q)}(z) \, \d z \, , \quad F_{L} = \sum_{q \in Q} F_{L}^{(q)} \, .
\end{equation}

Two conclusions are drawn from this investigation and exemplified in this section by presenting a single wind direction case; the findings are, however, generally true for other wind directions. First, because the flow is driven purely in the streamwise direction, the spanwise velocity is strongly influenced by local building geometry, causing the across-wind force on each building to alternate in sign. Consequently, when summed over all buildings (plane-averaged campus force $F_L/A_T$) or averaged over all wind directions for an individual building, the positive and negative spanwise contributions largely cancel each other, leaving a negligible net spanwise force. The average $F_L/A_T$ is indicated by the weak plane-averaged spanwise velocity in Fig.~\ref{fig: Fig2}(e).
Second, for an individual building under a specific wind direction, the across-wind force can be comparable in magnitude to its streamwise drag counterpart, as illustrated by the resultant force arrows in Fig.~\ref{fig: Fig4}(a) for $\theta = 180^\circ$: most resultant forces deviate noticeably from the wind direction due to the spanwise component. Fig.~\ref{fig: Fig4}(c) presents the probability distribution of the deviation angle of these resultant forces from the drag direction. 
The deviation angle ranges from $-90^\circ$ to $+90^\circ$, where positive values indicate clockwise deviation and negative values indicate anti-clockwise deviation. The probability distribution shows that most deviations are smaller than $\pm 45^\circ$ under this wind direction, indicating that the across-wind force does contribute, but the streamwise drag still dominates the resultant direction. Note that the peaks of the deviation angle distribution occur in the range $\pm (15^\circ-30^\circ)$ rather than near $0^\circ$, indicating that flow and force deflection are in fact a common occurrence for individual buildings on the campus. This is largely attributed to the building orientations, geometries and the wind-induced interference within the campus.

In conclusion, the across-wind force is of limited importance at the campus scale or when direction-averaged for an individual building, but can make a meaningful contribution to the resultant wind load on a specific building under a specific wind direction; this represents an interesting avenue for future work.
}

\subsection{Factors causing shielding effects} \label{sec: two parameters}
\begin{figure*}
            \centering
            \includegraphics[width=16cm]{ 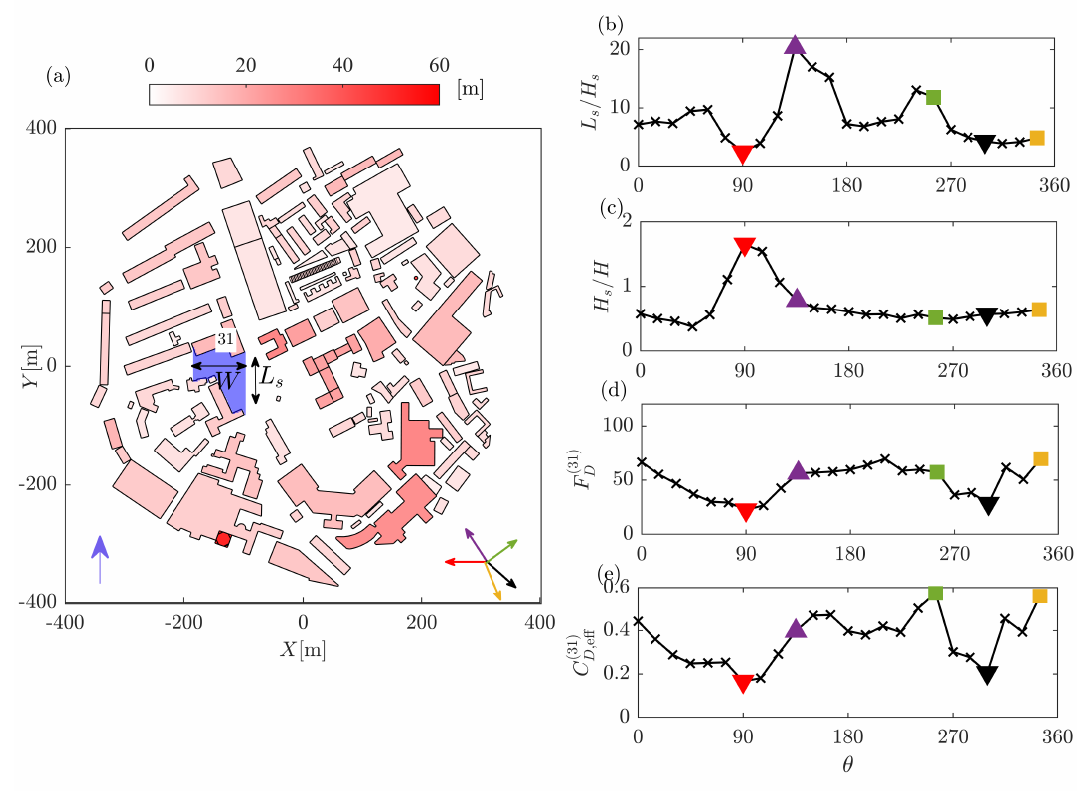}
            \caption{(a) Sketch of the open space (the purple patch) and shielding height in front of a target T-shaped building 31 under the wind direction $\theta = 180 \degree$ shown as the purple arrow. The horizontal double-headed arrow marks the maximum width of the fetch $W$ perpendicular to the wind direction, and the vertical double-headed arrow indicates the fetch length $L_s$ starting from the most windward point of the target building. Variation of the shielding effect parameters (b) $L_s/H_s$ and (c) $H_s/H$ of the target building with the wind direction. (d) Variation of cumulative integral drag $F_D^{(31)}$ and (e) drag coefficient $C_{D,\mathrm{eff}}^{(31)}$ of the target building with the wind direction. Several peaks and troughs in (b-e) are marked with their corresponding wind directions labelled in the bottom right of (a) in the corresponding colour, which are $\theta =  90\degree$ (red), $\theta = 135\degree$ (purple), $\theta = 255\degree$ (green), $\theta =  300\degree$ (black), $\theta = 345\degree$ (yellow), respectively.}
            \label{fig: Fig6}
\end{figure*}
The drag acting on a building varies with wind direction, due to two main factors: the change in frontal area exposed to the wind, and more importantly, the shielding effects induced by neighbouring buildings, which vary significantly with wind direction.

In a simple 2-building configuration, e.g., a target building and an upstream building located right in front of it, the drag on the target building is affected by both its spatial position relative to the upstream building and also its own height \citep{English1990, Lam2008}. For a fixed height of the target building, a shorter distance from the upstream building places the target building within its near-wake region, leading to stronger shielding and, consequently, a lower drag coefficient. Conversely, for a fixed spacing, increasing the height of the target building allows it to rise above the shielding region from the upstream building, resulting in a larger drag coefficient. These two mechanisms together influence the drag acting on individual buildings and underpin the strong directional dependence of drag observed across the campus.

However, in a realistic scenario, the situation is more complex, e.g., the shielding effect may arise from multiple neighbouring buildings. To systematically quantify the spatial relationships, two dimensionless parameters are introduced: 1) the relative open distance upstream of the target building, indicating its location relative to the wake flow, and 2) the relative height between the upstream and target buildings. Since the frontal area is inherently linked to building height, the second parameter also implicitly reflects variations in the frontal area of the building.

Figure \ref{fig: Fig6}(a) illustrates the definition of the first parameter. Taking a T-shaped building numbered 31 as an example, and assuming that the wind direction is $\theta = 180 \degree$ (shown at the bottom-left corner), the open space in front of this building, highlighted in purple, represents the region between the building and its nearest upstream building along the wind direction. Geometrically, this open space corresponds to the projection of the target building upstream along the wind direction until it intersects with other buildings. Denoting the plane area of the open space as $S$, the maximum width of the open space as $W$ (labelled by the horizontal double-headed arrow in Fig. \ref{fig: Fig6}a), and the mean height of the intersected upstream buildings as $H_s$, an averaged fetch distance is defined as $L_s = S/W$, which is the first parameter of interest, and is indicated by the vertical double-head arrow in Fig. \ref{fig: Fig6}(a). The second parameter, the relative building height, is defined as $H_s/H$, where $H$ is the target building height.

Figure \ref{fig: Fig6}(b, c) show the variations of the dimensionless parameters, the averaged fetch $L_s/H_s$ and relative building height $H_s/H$, with respect to the wind direction for target building 31. The parameter $L_s/H_s$ indicates the target building's position relative to the upstream wake flow: a larger $L_s/H_s$ indicates that the building is located further downstream from the upwind buildings. In contrast, the parameter $H_s/H$ reflects the degree of shielding provided by the upstream buildings: a higher $H_s/H$ corresponds to stronger shielding effects.
The wake parameter $L_s/H_s$ has a strong dependence on wind direction $\theta$ (Fig. \ref{fig: Fig6}b). A peak occurs at $\theta = 135\degree$ (marked in purple and also indicated by the same colour arrow in Fig. \ref{fig: Fig6}a), where a large open area exists upstream of the target building. As observed from the building layout, relatively lower values of $L_s/H_s$ are found when the wind comes to the target building from the first, second, and third quadrants, where the urban density is higher. It should be noted that $L_s/H_s$ also depends on the height of the upstream buildings $H_s$, although the variation of $H_s$ is not expected to be as large as that of the fetch $L_s$.
Figure \ref{fig: Fig6}(c) presents the height ratio parameter $H_s/H$. Since the height of the target building $H$ remains constant, this variation shows the change in the averaged upstream building height $H_s$. The value of $H_s$ remains relatively constant except for wind directions $60 \degree<\theta < 120\degree$, where it reaches a distinct peak at $\theta = 90\degree$ (marked in red and also indicated by the arrow in Fig. \ref{fig: Fig6}a), where the upstream buildings are relatively taller than their surroundings.

Figure \ref{fig: Fig6}(d, e) present the cumulative integral drag $F_D^{(31)}$ and the corresponding drag coefficient $C_{D,\mathrm{eff}}^{(31)}$ of the target building. Overall, both curves exhibit a similar shape across the wind directions. Two wind directions associated with the lowest drag are observed at $\theta = 90\degree$ and $\theta = 300\degree$ (highlighted in red and black in the figures, respectively). These directions generally correspond to local minima of $L_s/H_s$ (see Fig. \ref{fig: Fig6}b), consistent with the expectation that a shorter upstream fetch results in stronger shielding and, consequently, lower drag.
At $\theta = 300\degree$, the ratio $H_s/H \approx 0.55$, indicating a moderate level of shielding, whereas at $\theta = 90\degree$, $H_s/H$ has a maximum value (see Fig. \ref{fig: Fig6}c), strengthening the shielding and further reducing the drag. These observations are largely in line with the expectations from the effects of the two parameters. Conversely, two wind directions corresponding to the highest drag are found at $\theta = 255\degree$ (in green) and $\theta = 345\degree$ (in yellow). At $\theta = 255\degree$, $L_s/H_s$ approaches a local maximum while $H_s/H$ is relatively low. As the building height $H$ is fixed for the target building, this set of parameters indicates a low shielding height at the front and a relatively long-fetch location in the wake flow, both of which contribute to a high drag coefficient.

However, it is worth noting that 1) the two parameters introduced here do not always have a consistent effect for a given wind direction, and 2) although $L_s/H_s$ and $H_s/H$ capture the major directional effect, they do not fully describe the complex flow interactions. For example, the second drag coefficient peak at $\theta = 345\degree$ (in yellow) demonstrates this limitation: despite $L_s/H_s$ and $H_s/H$ being close to their values for the low-drag direction ($\theta = 300\degree$, in black), the drag at $\theta = 345\degree$ is unexpectedly large.
This could be because of the streamwise extent of this T-shaped building at $\theta = 345\degree$, where the vertical stem of the `T' is oriented almost perpendicular to this wind direction. This induces a larger local pressure difference between the windward and leeward sides, as the flow separation is likely to occur at this short distance.

\begin{figure*}
    \centering
    \includegraphics[width=15cm]{ 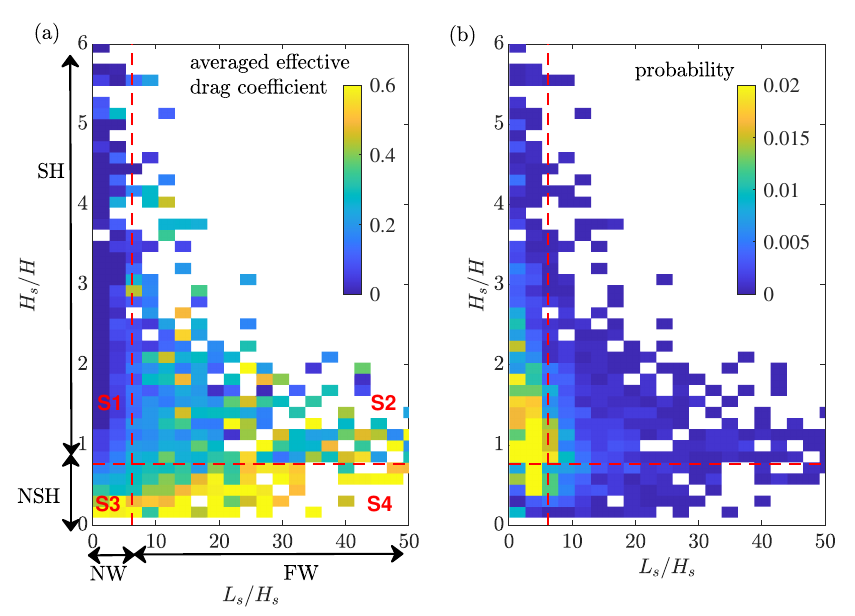}
    \caption{(a) Dependence of the bin-averaged drag coefficient of an individual building on the shielding parameters $L_s/H_s$ and $H_s/H$. (b) The probability of the shielding parameters $L_s/H_s$ and $H_s/H$ occurring on the campus. The red dashed lines classify the domain into four regimes (S1 -S4) according to the wake location and shielding situation.}
    \label{fig: Fig7}
\end{figure*}

To generalise the effect of the two dimensionless parameters on the buildings within the domain, we examine the drag coefficients of the 110 buildings over the 24 wind directions.
The values of the parameters, $L_s/H_s$ and $H_s/H$, are each divided into 50 bins. For every bin, the number of data points is counted to construct a probability with respect to the two parameters, reflecting the geometric characteristics of the building morphology. As shown in Fig. \ref{fig: Fig7}(b), most buildings on the campus are subject to wind conditions where $L_s/H_s < 10$ and $H_s/H < 2$. 
It is rare for a building to be situated in an extremely long wake or to be shielded by buildings that are substantially taller than itself. This spatial configuration, in turn, makes the campus model strongly heterogeneous with respect to wind direction.

Within each bin, the mean value of the individual building drag coefficients is calculated and shown in Fig. \ref{fig: Fig7}(a). Consistent with the previous analysis, for a given $H_s/H$, the drag coefficient tends to increase with increasing $L_s/H_s$, indicating that larger fetch leads to weaker shielding and higher drag. Conversely, for a fixed $L_s/H_s$, a higher shielding-height ratio $H_s/H$ corresponds to lower drag coefficients, confirming that taller upstream buildings provide stronger sheltering effects.

\begin{table*}
\centering
\renewcommand{\arraystretch}{1.2}
\begin{tabular*}{15cm}{@{\extracolsep{\fill}}lcccc}
\toprule
 & \multicolumn{2}{c}{Near-wake (NW)} & \multicolumn{2}{c}{Far-wake (FW)} \\
\cmidrule(lr){2-3} \cmidrule(lr){4-5}
 & Percentage & Average $C_{D,\mathrm{eff}}$ & Percentage & Average $C_{D,\mathrm{eff}}$\\
\midrule
Non-Shielding (NSH) & $12.5\%$ & $0.31$ & $9.4\%$ & $0.44$ \\
Shielding (SH)   & $54.9\%$ & $0.07$ & $23.2\%$ & $0.24$ \\
\bottomrule
\end{tabular*}
\caption{Probability of each regime (S1 to S4) occurring under all wind-direction cases, along with the average drag coefficient of the building for each situation.}
\label{tab: tab1}
\end{table*}

\review{
The parameter space can therefore be divided into four regimes using two thresholds: one on $L_s/H_s$ to separate near-wake (NW) from far-wake (FW) conditions, and another on $H_s/H$ to distinguish shielding (SH) from non-shielding (NSH) conditions. However, the determination of the thresholds is somewhat arbitrary and not unique in the literature; for example, the wake threshold value ranges from $L_s/H_s > 2$ to $L_s/H_s > 7$ in urban areas \citep{Belcher2012, SLR2012, Hertwig2019}, and the shielding threshold can physically-- and intuitively-- be set to $1$ by simply judging whether the upstream building is taller than the target building. Therefore, to identify an appropriate pair of thresholds that divide the drag coefficient into four regimes: (S1) near-wake shielding, (S2) far-wake shielding, (S3) near-wake non-shielding, and (S4) far-wake non-shielding, we apply a data-driven clustering approach to the full dataset of building drag coefficients. The optimal thresholds are selected to minimise the spread of drag coefficient values within each regime, while maximising the separation between different regimes. Details of the threshold-selection procedure and the corresponding sensitivity analysis are provided in Appendix~\ref{sec: threshold_selection}.

Following this approach, the optimal thresholds are identified as $L_s/H_s = 6.15$ and $H_s/H = 0.77$. This threshold pair, marked by the red dashed lines in Fig.~\ref{fig: Fig7}, produces a reasonably good collapse of the drag coefficient within each regime and a clear separation between the regime-wise distributions: S1 occupies the lowest drag-coefficient range, S4 the highest, while S2 and S3 form an intermediate band with limited overlap with S1 and S4. This reflects the combined effects of fetch and relative height: S1 corresponds to buildings that are both in the near wake and greatly shielded by taller upstream buildings, whereas S4 corresponds to buildings that are well exposed to the incoming flow with neither substantial wake nor shielding benefits. In S2 and S3, these two effects oppose one another and therefore partially cancel out, leading to intermediate drag coefficients. 

Statistics of the drag coefficient for each regime are summarised in Tab.~\ref{tab: tab1}. The averaged drag coefficient for each regime is consistent with the regime-wise distribution described above (and can also be seen in Fig.~\ref{fig: optimal_threshold}b in Appendix~\ref{sec: threshold_selection}). The probabilities of occurrence, also indicated in Fig.\ref{fig: Fig7}(b), reveal that more than half of the cases fall within the near-wake shielding regime (S1), while the far-wake non-shielding regime (S4) is the least common. The predominance of S1, S2 and S3 together reflects the generally dense campus layout, in which most buildings are surrounded by upstream buildings of comparable height.
}

\review{
\subsection{Evaluation of shielding factor}
\label{sec: excluding shielding effect}


\begin{figure}
    \centering
    \includegraphics{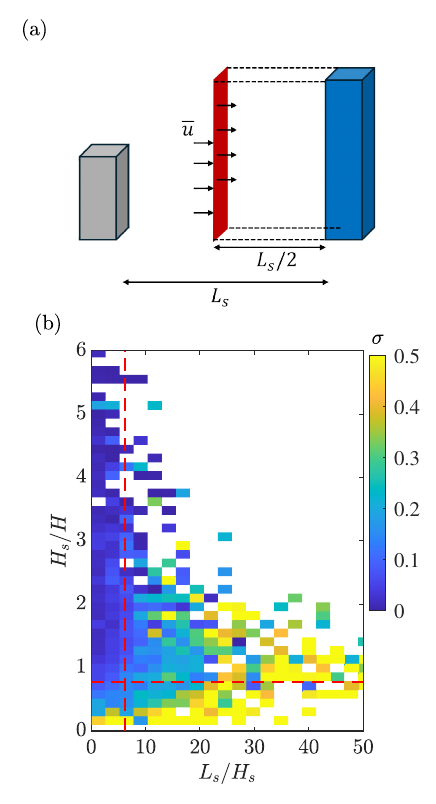}
    \caption{(a) Schematic of the effective velocity calculation, where the projected frontal area of the target building (blue cuboid) is defined at mid-fetch ($L_s/2$), and the effective velocity $\mathcal{U}$ is computed as the spatially averaged streamwise wind velocity $\overline{u}$ over this projection plane (the red plane). (b) Dependence of the bin-averaged exposure factor $\sigma$ of an individual building on the shielding parameters $L_s/H_s$ and $H_s/H$. The red dashed lines indicate the thresholds of regimes.}
    \label{fig: shielding factor}
\end{figure}
As previously introduced, the shielding factor is defined as the ratio of the building drag under shielding conditions to that in isolated conditions \citep{Khanduri1998}. However, a practical challenge in studying this within a realistic urban environment is that obtaining data, whether from numerical simulations or wind tunnel experiments, for all 110 buildings under isolated conditions is highly infeasible. Therefore, in this section, given that the building drag is largely governed by the approaching wind velocity, we alternatively assess the shielding factor directly by examining the averaged inflow wind velocity to the target building, and comparing it against the reference velocity. 

We determine a characteristic inflow velocity $\mathcal{U}$, which is defined as the spatially averaged streamwise velocity over the frontal projection plane of the target building at mid-fetch distance (see Figure~\ref{fig: shielding factor}a). This velocity directly characterises the reduced momentum flux experienced by the building due to upstream shielding. In terms of drag, this suggests a parameterisation as
\begin{equation}
  F_{D}^{(q)} = \frac{1}{2}\rho_0 A_{f}^{(q)} C_{D}^{(q)} {\mathcal{U}^{(q)}}^2 \, ,
\end{equation}
noting that, $C_{D}^{(q)}$ is the conventional `intrinsic' drag coefficient that represents the aerodynamic resistance of the building under equivalent non-shielded conditions. Compared with Eq.~\eqref{eq: F_D^q_e}, this equation incorporates the shielding effects into the velocity square term rather than into the effective drag coefficient.
Since both expressions ultimately describe the same drag force $F_{D}^{(q)}$, equating them yields:
\begin{equation}
  C_{D,\mathrm{eff}}^{(q)} = \sigma C_D^{(q)} \, , \quad \sigma = \left(\frac{\mathcal{U}^{(q)}}{U} \right)^2 \, .
\end{equation}
The parameter $\sigma$ can be regarded as a shielding factor. It quantifies the momentum retained despite the presence of upstream buildings, and therefore, reflects the degree to which the target building is shielded from the incoming flow relative to the reference condition. Neglecting other flow modification effects, such as channelling acceleration or wake-induced interference, a larger $\sigma$ represents less shielding from upstream buildings. Note that $U$ is a reference velocity, which is $U_{30}$ in this study.

Figure~\ref{fig: shielding factor}(b) shows the shielding factor of each building as a function of the geometric parameters $L_s/H_s$ and $H_s/H$. A clear trend is evident: buildings with a large fetch ratio $L_s/H_s$ and a low height ratio $H_s/H$ tend to have high $\sigma$, indicating very little shielding from upstream obstacles. Conversely, buildings with a small $L_s/H_s$ and a large $H_s/H$ exhibit low $\sigma$, corresponding to strong shielding by their upstream neighbours. This behaviour is consistent with the variation of $C_{D,\mathrm{eff}}^{(q)}$ observed in Fig.~\ref{fig: Fig7}(a), but $\sigma$ offers a more direct physical interpretation of shielding rather than embedding it in the effective drag coefficient. The shielding factor provides an explicit quantitative link between the effective drag coefficient $C_{D,\mathrm{eff}}^{(q)}$ and the intrinsic drag coefficient $C_D^{(q)}$.
}

\section{Conclusions} \label{sec: conclusions}

This study analysed 24 building-resolved large-eddy simulations over a realistic representation of the University of Bristol campus, comprising 110 buildings of varying shapes and orientations. The objective was to quantify how upstream shielding affects the building drag under various wind directions. At the campus scale, the approximately circular and relatively symmetric morphology leads to only moderate directional variation in the plane-averaged flow statistics and total drag. At the scale of individual buildings, however, the drag response is highly heterogeneous: approximately $80\%$ of the total building drag is produced by only $20\%$ of the buildings, and the drag coefficient of an individual building can vary substantially with wind direction.

The directional variation in individual-building drag is primarily controlled by shielding effects. Two geometric parameters were introduced to characterise this effect: the normalised upstream fetch, $L_s/H_s$, which describes the position of a target building relative to the wake of upstream buildings, and the relative shielding height, $H_s/H$, which measures the height of upstream obstacles relative to the target building. Larger $L_s/H_s$ generally corresponds to weaker wake shielding and higher drag, whereas larger $H_s/H$ corresponds to stronger shielding and lower drag. Using threshold values of $L_s/H_s = 6.15$ and $H_s/H = 0.77$, the building configurations were classified into four regimes. The near-wake shielding regime (S1) contains about half of all cases and is associated with negligible drag, whereas the far-wake non-shielding regime (S4), although less frequent, produces the largest drag coefficients and is therefore important for wind-loading assessment.

\review{
The study also highlighted that drag coefficients evaluated directly from the urban morphology should be interpreted as effective drag coefficients, because the computed drag forces already include the influence of shielding. To separate this shielding contribution from the intrinsic aerodynamic response, a shielding factor was defined from the approaching velocity at the target building. This factor provides a direct link between the effective drag coefficient and the conventional drag coefficient that would apply under equivalent non-shielded conditions but is not trivial to measure directly in an urban environment. Its variation with $L_s/H_s$ and $H_s/H$ is consistent with the effective drag coefficient, confirming that these two parameters capture the main geometric controls on shielding.

Most previous studies of shielding and interference effects have considered relatively idealised building configurations, whereas shielding in realistic urban environments remains much less explored. This study addresses this gap by analysing building drag and shielding in a dense, irregular campus morphology and by developing an LES-based study for quantifying these effects at the individual-building scale. Some quantitative findings, such as the exact drag distribution and occurrence probabilities, are specific to the Bristol campus case. However, the methodology itself, including the building-resolved drag evaluation, the geometric shielding parameters, the regime classification, and the shielding-factor approach, is general and can be applied to other complex urban environments.
}

The present work also opens several directions for future work. First, as discussed, the across-wind force can be important for individual buildings and should be examined further together with the resultant wind load. Second, the four shielding regimes provide a basis for morphological statistics that account explicitly for shielding. For example, \citet{Wong2010, Xu2022} suggested that frontal areas shielded by upstream buildings (so-called `effective frontal area') should be excluded when parameterising the drag coefficient. The regime classification developed here provides a practical way to identify such shielded areas in realistic urban morphologies.

\section*{Acknowledgments}
The authors gratefully acknowledge the support of the ARCHER2 UK National Supercomputing Service (project ARCHER2-eCSE05-3) and the NERC highlight grant ASSURE: Across-Scale ProcesseS in Urban Environment \\(NE/W002868/1, NE/W002841/1). The authors thank Dr Changchang Wang and the EnFlo group for providing the CAD file of the Bristol campus, and Dr James Matthews for his valuable local knowledge of the Bristol campus. The authors also thank reviewers for their valuable suggestions, which helped improve the quality of the manuscript.

\appendix
\review{
\section{Validation}\label{sec: validation}
Cross-code validation and grid-sensitivity tests are performed in this appendix. The code used for comparison is PALM 6.0 \citep{Maronga2020}, a widely used LES code that has been validated across many studies \citep{Gronemeier2021, Anders2023, Resler2021}. PALM 6.0 employs a 1.5-order subgrid-scale turbulence closure and a third-order Runge-Kutta fractional-step time integration scheme for the incompressible flow equations. The PALM simulations share the same settings as the current uDALES simulations: domain size $L_X \times L_Y \times L_Z = 800 \times 800 \times 300 \, \U{m^3}$, grid size  $N_X \times N_Y \times N_Z = 400 \times 400 \times 300$ (i.e., a grid spacing of $\Delta X \times \Delta Y \times \Delta Z = 2 \times 2 \times 1 \, \U{m^3}$), periodic (cyclic) boundary condition at lateral sides, and the same pressure-gradient forcing $\d P/\d x = 1.25 \times 10^{-5}$ kg m$^{-2}$s$^{-2}$ to drive the wind in the streamwise direction. A Courant-Friedrichs-Lewy (CFL) number of $0.9$ is used, resulting in a time step of approximately $1.00$ s. Simulations are conducted for four wind directions, $\theta = 0^\circ, 90^\circ, 180^\circ, 270^\circ$. Each simulation runs for $240\,000\,\U{s}$, of which the final $168\,000\,\U{s}$ are used to obtain converged time-averaged statistics. The time- and plane-averaged velocity profiles, as well as vertical profiles at four local stations, are collected and compared. The four local stations (A-D) are marked in Fig.~\ref{fig:bld_ids}: A at $(10 \U{m},10 \U{m})$, near the domain centre but on a building roof; B at $(300 \U{m}, 300 \U{m})$, in an open space away from buildings; C at $(50 \U{m}, -150\U{m})$, in a square surrounded by buildings; and D at $(-120 \U{m}, 150 \U{m})$, in a narrow street channel.

High-resolution uDALES simulations are conducted for the grid-sensitivity test using the same setting as the current simulation, but with the grid size doubled: $N_X \times N_Y \times N_Z = 800 \times 800 \times 600$, corresponding to a halved grid spacing of $\Delta X \times \Delta Y \times \Delta Z = 1 \times 1 \times 0.5 \, \U{m^3}$. The high-resolution simulations are conducted for two wind directions, $\theta = 0^\circ $, and $180^\circ$. The plane-averaged velocity profiles and local vertical profiles at the four stations A-D are, again, collected and compared. In addition, the streamwise drag forces on each individual building (buildings are numbered in Fig~\ref{fig:bld_ids}) are computed and compared between the two resolutions, enabled by the unique facet-based surface treatment implemented in uDALES (see Appendix~\ref{sec: volumetric density}).

\begin{figure}
    \centering
    \includegraphics[width=9cm]{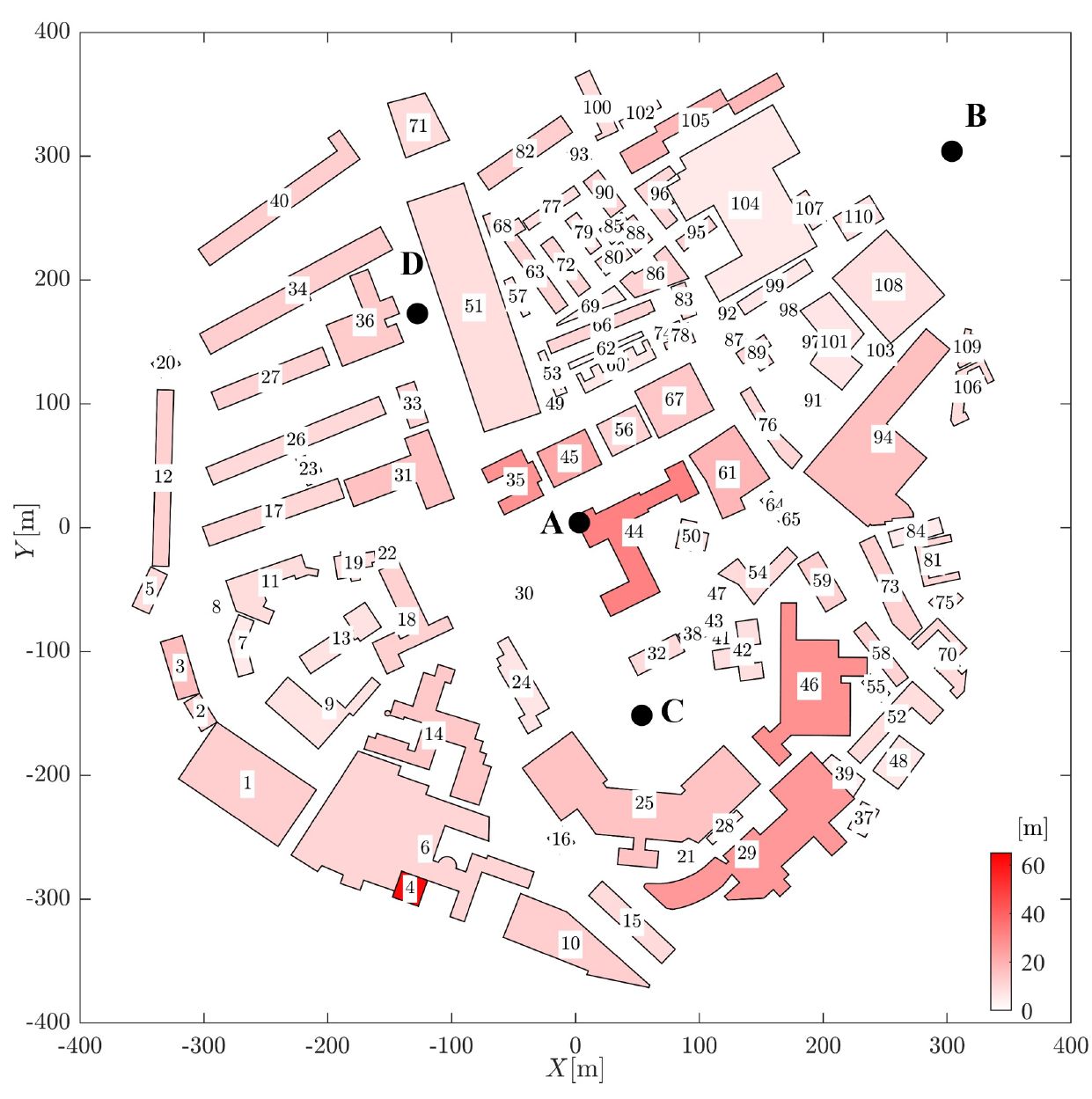}
    \caption{All 110 campus buildings, numbered from 1 to 110 in order from southwest to northeast, overlaid with four representative point stations (A-D) used for local velocity validation. The buildings are colored according to their height.}
    \label{fig:bld_ids}
\end{figure}

\begin{figure*}
    \centering
    \includegraphics[width=14.5cm]{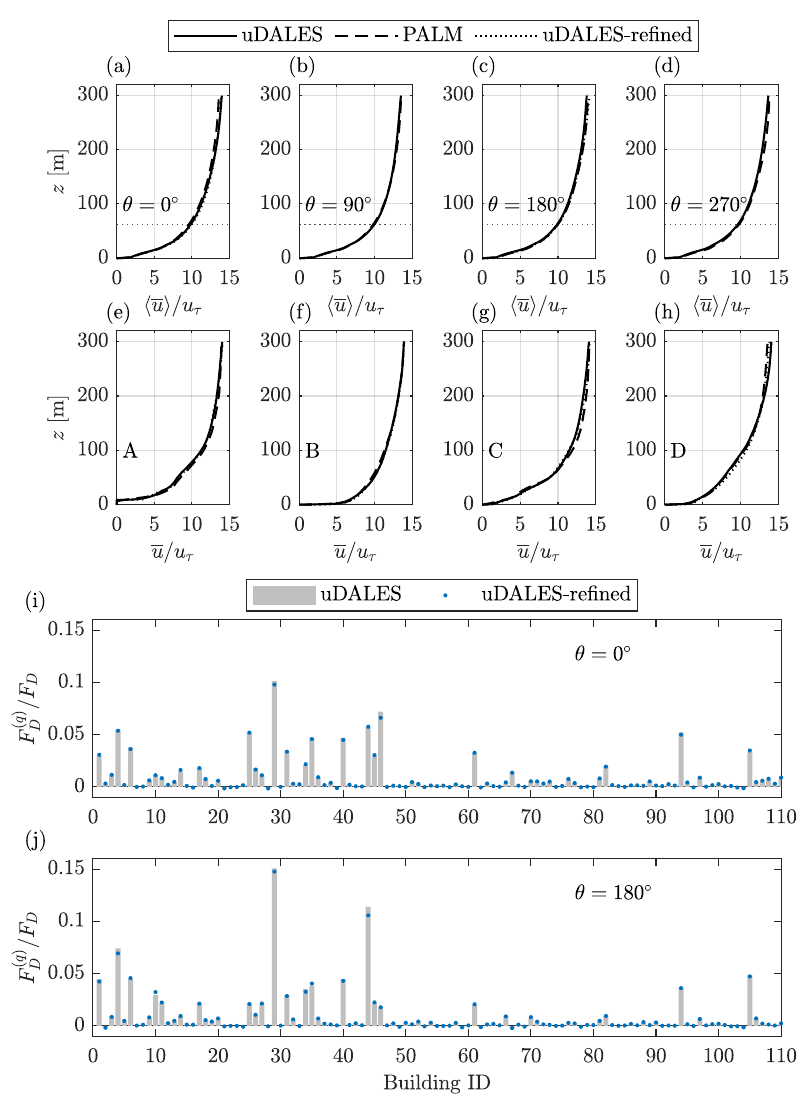}
    \caption{Grid-sensitivity analysis and cross-code verification. Results from PALM simulations at the current resolution and uDALES simulations at a doubled resolution (grid spacing halved) are compared against the current uDALES results. (a-d) Horizontally plane-averaged streamwise velocity profiles for four wind directions, $\theta = 0^\circ, 90^\circ, 180^\circ, 270^\circ$, respectively; refined uDALES results are available for $\theta = 0^\circ,  180^\circ$ only. The horizontal dotted line marks the maximum building height. (e-h) Local streamwise velocity profiles at four representative points A-D (marked in Fig.~\ref{fig:bld_ids}) for $\theta = 0^\circ$. Comparison of per-building drag force (normalised by the total streamwise drag force of all buildings) between the current-resolution and doubled-resolution uDALES simulations for wind directions (i) $\theta = 0^\circ$ and (j) $\theta = 180^\circ$.}
    \label{fig:validation_velocity}
\end{figure*}

Figure~\ref{fig:validation_velocity}(a-h) shows the comparison of velocity profiles among the three types of simulations.
Figures~\ref{fig:validation_velocity}(a-d) compare the time- and plane-averaged streamwise velocity profiles across the simulations; note that the high-resolution uDALES results are only available for $\theta = 0^\circ$ and $180^\circ$. The comparison shows good agreement across all simulations and wind directions, with $R^2$ values greater than $99\%$. The figures show the discrepancies are even smaller within the canopy region, indicated by the horizontal dotted line. Figures~\ref{fig:validation_velocity}(e-h) compare the local time-averaged velocity profiles at stations A-D for $\theta = 0^\circ$. The profiles again show excellent agreement with all $R^2$ values greater than $98.50\%$. Note that station A is located within a building footprint, so its velocity profile begins from the building roof level; all three simulations capture this, confirming that the building geometry is also handled consistently and correctly among the simulations, and that the results are well converged.

%
\label{tab: tab2}

Figures~\ref{fig:validation_velocity}(i, j) present the drag force on each individual building, calculated from the two uDALES simulations at different resolutions. All 110 campus buildings are numbered in Fig.~\ref{fig:bld_ids} in order from southwest to northeast. The drag is normalised by the total streamwise drag of all buildings. The results show good agreement between the two resolutions and across wind directions, with $R^2$ values of $99.76\%$ for $\theta = 0^\circ$ and $99.71\%$ for $\theta = 180^\circ$. The small discrepancies can only be observed for buildings with very large drag, confirming that the results are grid-converged.
}

\section{Calculation of distributed drag and frontal area} \label{sec: volumetric density}
\begin{figure*}
    \centering
    \includegraphics[width=15cm]{ 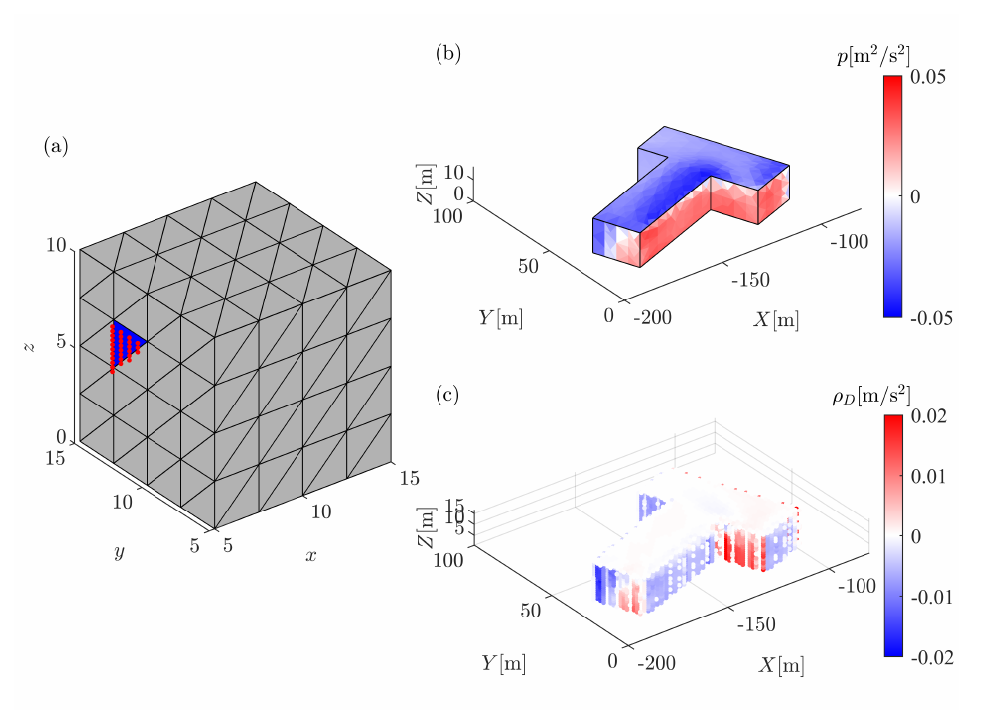}
    \caption{(a) Facet mesh of an example building, with a representative facet highlighted together with its associated nearest Cartesian grid cells. (b) The facet pressure of building 31 on the Bristol campus. (c) The drag density field $\rho_D$ converted from the facet data of building 31.}
    \label{fig:APP}
\end{figure*}
The uDALES code employs a surface mesh for building surfaces that is comprised of triangular facets. Each facet is associated with a set of Cartesian boundary cells as shown in Fig. \ref{fig:APP}(a). These are the cells immediately adjacent to the facet that are used to apply the momentum and scalar exchanges between the surface and the fluid \citep[see][for more details]{Owens2024}. However, this mapping is also extremely useful for diagnostic purposes, such as calculating vertically distributed quantities, frontal areas and forces on individual buildings. 

The conservation of an arbitrary surface quantity $\phi$ requires that \citep{MVR2025}:
\begin{equation} \label{eq:Phi_con}
  \int \oint_{\partial \Omega_f(z)} \frac{\phi}{|\vec N_\perp|} \d s \d z = \oiint_{\partial \Omega_f} \phi \d S,
\end{equation}
which illustrates that integrating the line integral over $z$ is identical to the total flux exchanged between the solid surface and the fluid. By discretising this identity, it becomes clear how to deal with the surface normal since $\Delta s_m/|\vec N_\perp|_m = A_m/\Delta z_{K_m}$ \cite{MVR2025}. This also shows that the singularity that emerges if the surface is horizontal is only apparent -- a finite amount of exchange occurs over an infinitesimal increase in $z$.

uDALES defines a volumetric density $\rho_\phi$ that contains the boundary exchanges and which is only non-zero in the cells next to the boundary \citep{MVR2025}:
\begin{equation}
  \label{eq:Phi}
  \rho_{\phi;ijk} = \sum_{m \in M_{ijk}} \frac{\phi_m A_m}{\Delta x \Delta y \Delta z_{K_m}} \, ,
\end{equation}
where each cell-facet $m$ has area $A_m$ and contribution $\phi_m$. The term $M_{ijk} \in \{m: I_m = i, J_m = j, K_m = k \}$ is the set of all cell-facets associated with cell $i,j,k$. 
From \eqref{eq:Phi}, it is clear that the volumetric density is a distribution comprising a large number of discrete delta functions. This makes its local value resolution-dependent and not directly meaningful. However, its integral is well defined. By substituting \eqref{eq:Phi} into the discrete form of \eqref{eq:Phi_con}, it is straightforward to show that
\begin{equation}
  \label{eq:PhiA}
  \sum_{i,j,k} \rho_{\phi;ijk}\Delta x \Delta y \Delta z_k \, = \sum_m \phi_m A_m,
\end{equation}
The left-hand side is a discrete volume integral, which we can rewrite as $\int \rho_\phi \d V = A_T \langle \rho_\phi \rangle$ using \eqref{eq: plav}. This implies that 
\begin{equation}
    \sav{\rho_\phi}(z) = \frac{1}{A_T} \oint_{\partial \Omega_f(z)} \frac{\phi}{|\vec N_\perp|} \d s \, .
\end{equation}
This is an important result, since it links the superficial average of the volumetric density directly to the surface integral. Indeed, using $\phi_m = -(\overline p_m \vec e_x  - \nu (\nabla \overline u)_m) \cdot \vec N_m$, we obtain the distributed drag forcing
$f_D(z) = \sav{\rho_D}$ where $\rho_D$ is the volumetric drag density. It is straightforward to obtain the individual contributions of skin and form drag using this approach.

Figure \ref{fig:APP}(b) shows the facet pressure $p_m$ on one building surface, and Fig. \ref{fig:APP}(c) shows the corresponding volumetric drag density $\rho_D$ distributed over the facet-associated Cartesian grid cells. Besides $p_m$, the volumetric drag density $\rho_D$ also incorporates the shear stress contribution, which is also a surface property (although it is not expected to be as significant as the pressure contribution). 

One caveat for the approach used is that one has to make sure that facets are sufficiently small to capture spatial variations in surface pressure. Indeed, inside uDALES, the pressure on a facet is obtained by taking the area-weighted average of the pressure inside all boundary cells associated with the facet. Any spatial variations across the surface are averaged out, which is why it is advisable to generate facets that are no larger than a few cells if surface properties are of interest.

Another useful application of this method is in the calculation of the frontal area for the entire urban surface. By setting $\phi_m = - \min(\vec e_u \cdot \vec N_m,0)$ where $\vec e_u$ is the unit vector associated with the wind direction, the quantity $\phi_m A_m$ represents the contribution of facet $m$ to the frontal area. Denoting the associated volumetric frontal area density, $\rho_{A_f}$, the frontal area is simply $A_f = \int \rho_{A_f} \d V$; see Eq.~\eqref{eq:PhiA}.

\section{Principal component analysis of building geometry} \label{sec: WPCA}
\begin{figure}
    \centering
    \includegraphics[width=7cm]{ 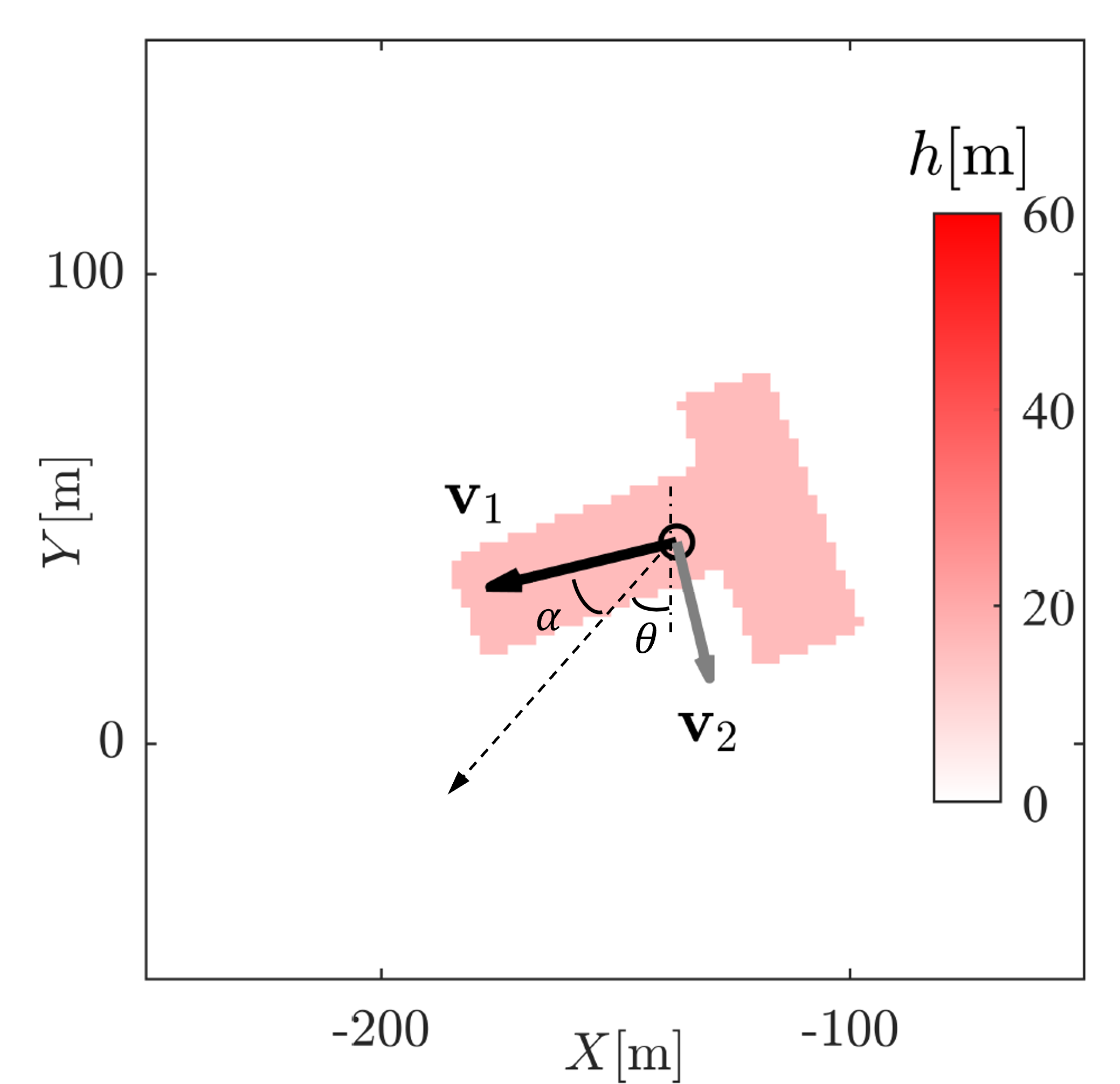}
    \caption{The height-masking footprint $h$ of building 31, overlaid with its principal ($\vec v_1$ in black) and secondary ($\vec v_2$ in grey) directions, originating at the building centroid (indicated by a circle). The dashed arrow indicates the wind direction $\theta$, while $\alpha$ denotes the angle between the wind direction and $\vec v_1$. }
    \label{fig:APP2}
\end{figure}
This section describes the method for obtaining the principal and secondary directions of an urban geometry. This method is well-known for porous media flows \citep{Athanatopoulou2008, Jolliffe2016}. The novelty of the approach described here is that the height distribution of the buildings is taken into account.  Let $h(\vec x_h)$ be the height function for the urban surface, where $\vec x_h = [x, y]^T$ and where $h$ is the local building height if $\vec x_h$ is inside a building and is 0 otherwise. 

The height-weighted building centroid $\vec x_c$ is given by
\begin{equation}
    \vec x_c = \frac{\int_\Omega h(\vec x_h) \vec x_h \, \d \vec x_h}{\int_\Omega h(\vec x_h) \vec \, \d \vec x_h} \, 
\end{equation}
where $\Omega$ is the horizontal extent of the urban surface.
A $2 \times 2$ weighted covariance matrix $C$ is constructed to characterise the distribution tensor of the structure
\begin{equation}
    C = \frac{\int_\Omega h(\vec x_h) (\vec x_h - \vec x_c) (\vec x_h - \vec x_c)^T \, \d \vec x_h}
             {\int_\Omega h(\vec x_h) \vec \, \d \vec x_h}\, ,
\end{equation}
which is decomposed into its eigenvectors and eigenvalues using the identity $C V = V \Lambda$, where  $V = [\vec v_1, \vec v_2]$ contains the eigenvectors and $\Lambda$ the eigenvalues. The primary eigenvector $\vec v_1$ aligns with the principal direction of the building, while the secondary eigenvector $\vec v_2$ aligns with the secondary direction. The length of the eigenvectors also indicates the characteristic length scale in the corresponding direction.
The angle between the wind direction and $\vec v_1$ (denoted by $\alpha$), therefore, characterises the relative alignment between the incoming wind and the principal direction of the building geometry.
This method can be applied to either a group of buildings or an individual building, as it operates on the height field and characterises the spatial distribution of the height-weighted geometry.

Applying the above approach, as an example, vectors $\vec v_1$ and $\vec v_2$ in Fig. \ref{fig:APP2} show the principal and secondary directions of building 31, respectively, as well as the characteristic length of extension in both directions. 
\review{

\section{Selection of regime thresholds} \label{sec: threshold_selection}
\begin{figure*}
    \centering
    \includegraphics[width=15cm]{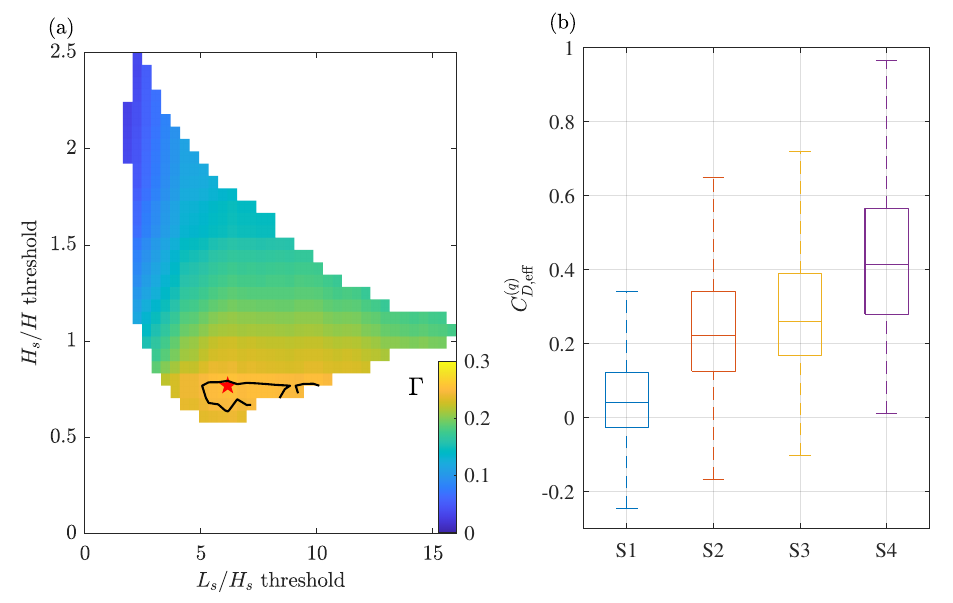}
    \caption{(a) The colour plot of the separation parameter $\Gamma$ against different threshold pairs $L_s/H_s , H_s/H $ . The larger the $\Gamma$, the better the thresholds satisfy the separation requirements. The star marks the optimal thresholds, and the black contour marks $\Gamma = 0.25$. (b) Box plots of the building drag coefficient for the four regimes identified by the optimal threshold. S1: shielding near-wake, S2: shielding far-wake, S3: non-shielding near-wake, S4: non-shielding far-wake.}    
    \label{fig: optimal_threshold}
\end{figure*}

As described in \S~\ref{sec: two parameters}, the fetch ratio $L_s/H_s$ and height ratio $H_s/H$ thresholds are determined using a data-driven clustering analysis applied to the full dataset of building drag coefficients. The objective is to identify a threshold pair that partitions the parameter space into four regimes while yielding drag-coefficient distributions that are as compact as possible within each regime, and as distinct as possible between regimes.

To quantify these two requirements, a separation parameter $\Gamma$ is defined as the ratio of between-regime variance $V_b$ (corresponding to the second criterion above) to within-regime variance $V_w$ (corresponding to the first criterion) of the drag coefficient $C_{D,\mathrm{eff}}^{(q)}$ across the four regimes, i.e.,
\begin{equation}
    \Gamma = \frac{V_b}{V_w} \, .
\end{equation}
Here, the between-regime variance $V_b$ represents the separation of the drag coefficient between the four regimes:
\begin{equation}
    V_b = \sum_{k=1}^{4} n_k \left( \overline{C_{D,\mathrm{eff};k}^{(q)}} - \overline{C_{D,\mathrm{eff}}^{(q)}} \right)^2 \, ,
\end{equation}
where $k$ denotes the regime index (from $1$ to $4$), $n_k$ is the number of data samples in regime $S_k$, $\overline{C_{D,\mathrm{eff}}^{(q)}}$ is the overall average of all drag coefficient samples, and $\overline{C_{D,\mathrm{eff};k}^{(q)}}$ is the average drag coefficient of regime $S_k$.
The within-group variance $V_w$ represents the spread of the drag coefficient within each regime:
\begin{equation}
    V_w = \sum_{k=1}^{4} \sum_{i \in S_k} \left( C_{D,\mathrm{eff}}^{(q)} - \overline{C_{D,\mathrm{eff};k}^{(q)}} \right)^2 \, .
\end{equation}
By definition, a larger $\Gamma$ indicates that the selected thresholds better satisfy both requirements simultaneously. To ensure that the resulting regimes are practically meaningful, an additional constraint is applied to the threshold selection: each regime must contain at least $5\%$ of the total data samples; otherwise, $\Gamma$ is set to NaN for that threshold pair.

Figure~\ref{fig: optimal_threshold}(a) shows the $\Gamma$ values against different threshold pairs, where the maximum value $\Gamma = 0.27$, i.e., the optimal threshold, is marked at $L_s/H_s = 6.15$ and $H_s/H = 0.77$. The corresponding box plot of the building drag coefficient for each regime obtained using the optimal thresholds is presented in Fig.~\ref{fig: optimal_threshold}(b). The box plot shows that the near-wake shielding regime (S1) yields a median drag coefficient of only approximately $0.04$, as both the near-wake location and large height ratio act to suppress drag. In contrast, the far-wake non-shielding regime (S4) has the highest drag coefficients, with a median of approximately $0.41$, as buildings are highly exposed to the incoming wind with neither wake nor shielding protection. The remaining two regimes, far-wake shielding (S2) and near-wake non-shielding (S3), exhibit intermediate and comparable drag coefficients, because in each case the two factors act in opposing directions and partially cancel, resulting in similar distributions for S2 and S3. The box plot reveals a clear separation, with S1 occupying the lowest coefficient range while S4 occupies the highest, and S2 and S3 together form an intermediate band with limited overlap with S4, confirming that the selected thresholds are physically meaningful and that the clustering is appropriate.

The sensitivity of the threshold choice is also illustrated in Fig.~\ref{fig: optimal_threshold}(a). The black contour encloses a region of threshold pairs that yield similarly high $\Gamma$ values (greater than $0.25$) as the selected optimal pair; however, note that this region is small relative to the full parameter space shown in Fig.~\ref{fig: Fig7}(a). Within this region, $\Gamma$ is relatively insensitive to fetch threshold $L_s/H_s$ but more sensitive to the height threshold $H_s/H$. This is primarily attributable to the distribution of data samples shown in Fig.~\ref{fig: Fig7}(b): in this contour region, shifting the $H_s/H$ threshold redistributes a larger number of samples across the regimes than shifting the $L_s/H_s$ threshold. Nevertheless, as reflected by $\Gamma$, varying the thresholds within this contour region does not substantially degrade the clustering quality, although it does shift the statistics shown in the box plots; for instance, increasing the fetch ratio $L_s/H_s$ threshold would marginally raise the median values of all the regimes simultaneously. Overall, these results support the threshold pair adopted in this work.
}

\bibliographystyle{cas-model2-names}

\bibliography{References.bib}

\end{document}